\documentclass[floatfix,twocolumn,aps,pra,10pt,amsmath,amssymb,superscriptaddress,notitlepage,longbibliography]{revtex4-2}
\usepackage{silence}
\usepackage[T1]{fontenc}
\usepackage{comment}
\usepackage{textcomp}
\usepackage{graphicx}
\usepackage[svgnames]{xcolor}
\usepackage{hyperref}
\hypersetup{colorlinks,citecolor=DarkGreen,linkcolor=FireBrick,urlcolor=FireBrick,linktocpage,unicode}
\usepackage[normalem]{ulem}
\usepackage{braket}
\usepackage{bm}
\usepackage{here}
\usepackage{cancel}

\allowdisplaybreaks[2]
\usepackage[pagewise,mathlines]{lineno}

\DeclareRobustCommand{\TAdel}{\bgroup\markoverwith{\textcolor{Green}{\rule[.5ex]{2pt}{1pt}}}\ULon}

\DeclareRobustCommand{\SMdel}{\bgroup\markoverwith{\textcolor{Red}{\rule[.5ex]{2pt}{0.7pt}}}\ULon}
\begin{document}
\title{High-speed \texorpdfstring{$R_{ZZ}$}{RZZ} gates in Kerr-cat qubits via fast-forward scaling}
\author{Takaaki Aoki}
\affiliation{Global Research and Development Center for Business by Quantum-AI Technology (G-QuAT), National Institute of Advanced Industrial Science and Technology (AIST), 1-1-1 Umezono, Tsukuba, Ibaraki 305-8568, Japan}
\author{Shumpei Masuda}
\email{shumpei.masuda@aist.go.jp}
\affiliation{Global Research and Development Center for Business by Quantum-AI Technology (G-QuAT), National Institute of Advanced Industrial Science and Technology (AIST), 1-1-1 Umezono, Tsukuba, Ibaraki 305-8568, Japan}
\affiliation{NEC-AIST Quantum Technology Cooperative Research Laboratory, National Institute of Advanced Industrial Science and Technology (AIST), 1-1-1 Umezono, Tsukuba, Ibaraki 305-8568, Japan}
\begin{abstract}
Kerr parametric oscillators can encode Kerr-cat qubits, whose biased-noise nature can reduce the hardware overhead required for fault-tolerant quantum computation. In a recently proposed $R_{ZZ}$-gate scheme for Kerr-cat qubits, a tunable coupler must be displaced while avoiding nonadiabatic excitation. To address this challenge, we propose and theoretically analyze a protocol that applies fast-forward scaling to the coupler displacement. The protocol suppresses nonadiabatic excitation of the coupler and enables high speed $R_{ZZ}$ gates.
\end{abstract}
\maketitle

\section{Introduction}
Kerr parametric oscillators (KPOs), i.e., parametrically driven oscillators with Kerr nonlinearity~\cite{Cochrane1999,Goto2016,PhysRevA.93.050301,Puri2017}, have been extensively studied in recent years.
These systems support two coherent states with opposite phases, which can be used to encode a biased-noise qubit whose bit-flip rate is much smaller than the phase-flip rate. 
Such a qubit is referred to as a Kerr-cat qubit or a KPO qubit~\cite{PhysRevX.9.041009,Grimm2020,PhysRevApplied.21.014030,PhysRevA.110.012463,Iyama2024,PhysRevX.14.031040,PhysRevX.14.041049,Hoshi2025,Ding2025,c661-yr2z,fyft-6kqg,PhysRevApplied.23.034060,doi:10.1073/pnas.2520479123,dryl-275b}.
The biased noise can reduce hardware overhead for fault-tolerant quantum computing~\cite{doi:10.1126/sciadv.aay5901,PhysRevResearch.4.013082,PRXQuantum.2.030345}, making Kerr-cat qubits a promising platform for scalable quantum computation.

To achieve faster gate operations in Kerr-cat qubits while maintaining high fidelity, shortcuts to adiabaticity (STA) techniques~\cite{doi:10.1002/9781119096276.ch3,TORRONTEGUI2013117,delCampo_2019,RevModPhys.91.045001,Hatomura_2024} such as counterdiabatic (CD) driving~\cite{PhysRevLett.85.1626,Demirplak2003,Demirplak2005,Berry_2009,doi:10.1098/rsta.2021.0272} and its multi-transition variant, derivative removal by adiabatic gate (DRAG)~\cite{PhysRevLett.103.110501,PhysRevA.83.012308,PhysRevA.88.062318,Theis_2018}, have been employed~\cite{PhysRevResearch.4.013082,PhysRevApplied.18.034076,PhysRevResearch.6.013192,10.1063/5.0241996}. 
STA is a class of protocols that reproduce the final outcome of adiabatic evolution, which is intrinsically stable but suffers from decoherence due to its slowness~\cite{SCHLOSSHAUER20191}, within a much shorter time.

A typical entangling gate for Kerr-cat qubits is a $ZZ$ rotation ($R_{ZZ}$ gate), which is based on a $ZZ$ coupling~\cite{fors2024comprehensiveexplanationzzcoupling}. Residual $ZZ$ couplings can cause correlated and nonlocal crosstalk errors~\cite{Sarovar2020detectingcrosstalk}.
To suppress such errors, we previously proposed a tunable-coupler-based $ZZ$-coupling scheme for
Kerr-cat qubits~\cite{10.1063/5.0241315}.
In this scheme, the $ZZ$ interaction is controlled by varying the coupler frequency in time, thereby
implementing an $R_{ZZ}$ gate.
However, sufficiently slow modulation is required to avoid nonadiabatic excitation of the coupler.

In this work, we utilize fast-forward scaling theory (FFST) to accelerate the $R_{ZZ}$ gate while
suppressing nonadiabatic transitions.
FFST exploits a nontrivial scaling property of quantum dynamics and provides a systematic method for designing control parameters that realize speed-controlled evolution
~\cite{PhysRevA.78.062108,doi:10.1002/9781119096276.ch3,doi:10.1098/rsta.2021.0278}.

In particular, FFST enables acceleration, deceleration, and even time-reversal of quantum evolution. The theory was further extended to accelerate quantum adiabatic dynamics~\cite{doi:10.1098/rspa.2009.0446}, which is a form of STA.
Over the past two decades, FFST applied to adiabatic dynamics has been investigated in a wide variety of systems, including cold atoms~\cite{doi:10.1098/rspa.2009.0446,PhysRevA.86.013601,PhysRevA.87.033630,PhysRevLett.113.063003,PhysRevA.94.063418,PhysRevA.103.023307}, charged particles~\cite{PhysRevA.84.043434,Kiely_2015,Masuda2015,An2016}, many-body systems~\cite{PhysRevA.86.063624}, spin systems~\cite{PhysRevA.96.052106,PhysRevA.99.062116}, discrete systems~\cite{PhysRevA.89.033621,PhysRevA.89.042113,10.21468/SciPostPhys.15.1.036}, and even relativistic quantum systems~\cite{Deffner_2016,Sugihakim_2021}.
Furthermore, several extensions of FFST have been developed to broaden its applicability~\cite{Patra_2017,PhysRevResearch.3.013087,Masuda2022}.
To our knowledge, however, FFST has not previously been applied to the tunable-coupler-based $R_{ZZ}$ gate considered here.

In the tunable-coupler-based $R_{ZZ}$ gate, the gate speed is limited by the requirement that the coupler follow its instantaneous coherent state adiabatically. To overcome this limitation, we apply fast-forward scaling theory to the displacement of the coupler. The coupler dynamics can be mapped onto those of an off-resonantly driven resonator, which enables the construction of a fast-forward protocol through detuning modulation.
The remainder of this paper is organized as follows. In Sec.~\ref{sec:System}, we introduce the KPO-coupler system, review the adiabatic $R_{ZZ}$-gate protocol, and derive the effective coupler model that motivates the subsequent development.
Before presenting the resulting fast-forwarded $R_{ZZ}$ gate in Sec.~\ref{sec:Application}, we introduce three preliminary sections (Secs.~\ref{sec:Review}--\ref{sec:Delta}).
In Sec.~\ref{sec:Review}, we review the fast-forward approach for realizing perfect displacement of a harmonic oscillator through modulation of the potential minimum.
In Sec.~\ref{sec:Omega}, we propose a scheme for perfect displacement of a superconducting resonator
by modulating the amplitude of an off-resonant drive, based on the review in Sec.~\ref{sec:Review}.
In Sec.~\ref{sec:Delta}, we study the time-scaling property of the superconducting resonator and
combine the results with those obtained in Sec.~\ref{sec:Omega} to realize perfect
displacement through detuning modulation. In Sec.~\ref{sec:Application}, we apply the method
to the coupler and realize a fast-forwarded $R_{ZZ}$ gate.
Finally, in Sec.~\ref{sec:Conclusions}, we present our conclusions.

\section{System and adiabatic \texorpdfstring{$R_{ZZ}$}{RZZ} gate}
\label{sec:System}
\begin{figure}
    \centering
    \includegraphics[width=0.33\textwidth]{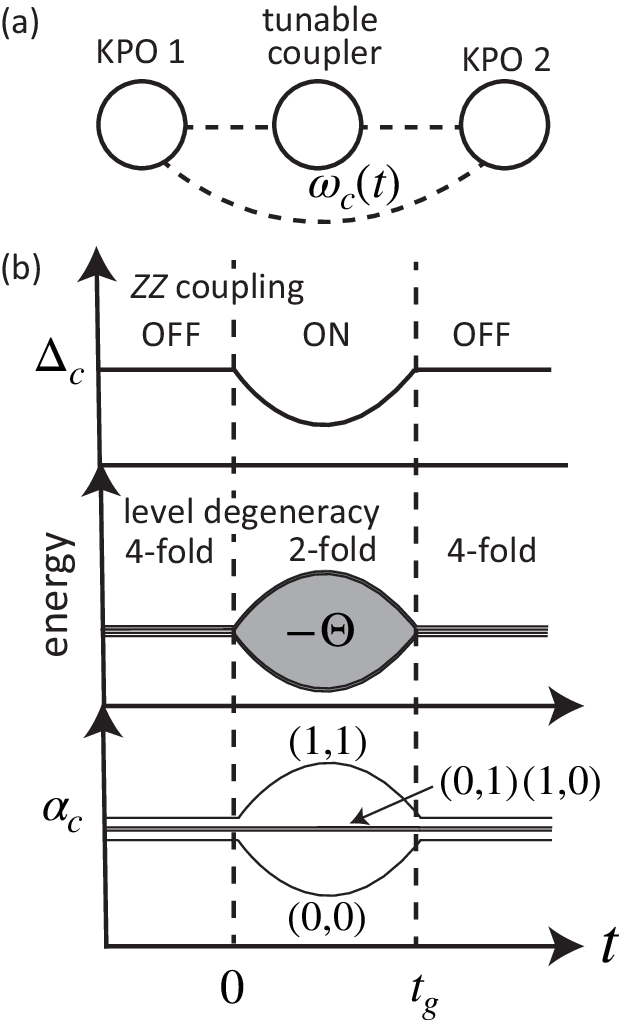}
    \caption{(a) Schematic of a system consisting of a frequency-tunable resonator (coupler, subsystem c) and two Kerr parametric oscillators (KPOs, subsystems $1$ and $2$).
    (b) Mechanism of the $ZZ$ coupling between Kerr-cat qubits in the adiabatic regime.
    Top: the coupler's detuning, $\Delta_{\mathrm{c}}(t)$.
    Middle: the four eigenenergies of the system in the first order of perturbation in Eqs.~\eqref{eq:E00E11} and \eqref{eq:E01E10} corresponding to the four states in Eqs.~\eqref{eq:psi00t}--\eqref{eq:psi11t}.
    $\Theta$ is the rotation angle of the $R_{ZZ}$ gate.
    Bottom: the amplitudes of the four coherent states of the coupler, $\pm\alpha_{\mathrm{c}}^{+}(t)$ and $\pm\alpha_{\mathrm{c}}^{-}(t)$.
    In the figure, $\alpha_{\mathrm{c}}^{-}(t)=0$, while in Ref.~\cite{10.1063/5.0241315}, $\alpha_{\mathrm{c}}^{-}(t)\neq0$.
    $t_{\mathrm{g}}$ is the gate time.
    \label{fig:2KPO1coupler}}
\end{figure}

We consider a system consisting of a frequency-tunable resonator (coupler, subsystem c)
and two Kerr parametric oscillators (KPOs, subsystems 1 and 2),
as shown in Fig.~\ref{fig:2KPO1coupler}(a).
The KPOs encode Kerr-cat qubits and are parametrically pumped at frequency $\omega_p$, while the tunable coupler mediates a controllable interaction between them. 

The Hamiltonian of the system in a frame rotating at frequency $\omega_p/2$
under the rotating-wave approximation is given by~\cite{10.1063/5.0241315}
\begin{linenomath}
\begin{align}
    \hat{H}(t)
    &=\sum_{j=1,2}\hat{H}_j(t)
    +\hat{H}_{\mathrm c}(t)
    +\hat{H}_{\mathrm I},
    \label{eq:Hamiltonian_totalR}
    \\
    \hat{H}_j(t)/\hbar
    &=
    -\frac{K_j}{2}\hat{a}_j^{\dagger2}\hat{a}_j^2
    +\frac{p_j}{2}
    \left(
        \hat{a}_j^{\dagger2}
        +\hat{a}_j^2
    \right)
    +\Delta_j(t)\hat{a}_j^\dagger\hat{a}_j,
    \label{eq:Hamiltonian_jR}
    \\
    \hat{H}_{\mathrm c}(t)/\hbar
    &=
    \Delta_{\mathrm c}(t)
    \hat{a}_{\mathrm c}^{\dagger}
    \hat{a}_{\mathrm c},
    \label{eq:Hamiltonian_cR}
    \\
    \hat{H}_{\mathrm I}/\hbar
    &=
    \sum_{j=1,2}
    g_{j\mathrm c}
    \left(
        \hat{a}_j^\dagger\hat{a}_{\mathrm c}
        +
        \hat{a}_j\hat{a}_{\mathrm c}^\dagger
    \right)
    +
    g_{12}
    \left(
        \hat{a}_1^\dagger\hat{a}_2
        +
        \hat{a}_1\hat{a}_2^\dagger
    \right).
\end{align}
\end{linenomath}
Here, $K_j$ is the Kerr nonlinearity of the $j$th KPO,
$p_j$ is the parametric-drive amplitude,
$\Delta_\lambda=\omega_\lambda-\omega_p/2$
($\lambda\in\{1,2,\mathrm c\}$) is the detuning from half the pump frequency,
and $g_{j\mathrm c}$ and $g_{12}$ are coupling strengths.
We assume that the coupler has negligible Kerr nonlinearity.

The Hamiltonian in Eq.~\eqref{eq:Hamiltonian_totalR} can be rewritten as~\cite{10.1063/5.0241315}
\begin{linenomath}
\begin{align}
    \hat{H}(t)&=\hat{H}_{0\mathrm{th}}(t)+\hat{H}_{ZZ}(t)+\sum_{j=1,2}\hat{H}_{X_j}(t), \\
    \hat{H}_{0\mathrm{th}}(t)/\hbar&=\sum_{j=1,2}\left[-\frac{K_j}{2}(\hat{a}_j^{\dag2}-\alpha_j^2)(\hat{a}_j^{2}-\alpha_j^2)+\frac{K_j\alpha_j^4}{2}\right] \notag \\
    &\quad\mbox{}+\Delta_{\mathrm{c}}(t)\left(\hat{a}_{\mathrm{c}}^{\dag}+\frac{g_{1\mathrm{c}}}{\Delta_{\mathrm{c}}(t)}\hat{a}_1^{\dag}+\frac{g_{2\mathrm{c}}}{\Delta_{\mathrm{c}}(t)}\hat{a}_2^{\dag}\right) \notag \\
    &\quad\times\left(\hat{a}_{\mathrm{c}}+\frac{g_{1\mathrm{c}}}{\Delta_{\mathrm{c}}(t)}\hat{a}_1+\frac{g_{2\mathrm{c}}}{\Delta_{\mathrm{c}}(t)}\hat{a}_2\right), \\
\hat{H}_{ZZ}(t)/\hbar &= \left(g_{12}- \frac{g_{1\mathrm c}g_{2\mathrm c}} {\Delta_{\mathrm c}(t)} \right)
\left(\hat{a}_1^\dagger\hat{a}_2+\hat{a}_1\hat{a}_2^\dagger\right),
    \label{eq:HZZt} \\
    \hat{H}_{X_j}(t)/\hbar&=\left(\Delta_j(t)-\frac{g_{j\mathrm{c}}^2}{\Delta_{\mathrm{c}}(t)}\right)\hat{a}_j^{\dag}\hat{a}_j,
\end{align}
\end{linenomath}
where $\hat H_{ZZ}(t)$ is the term responsible for the effective $ZZ$ coupling between the KPOs,
while $\hat H_{X_j}(t)$ induces an unwanted X-axis rotation
($R_X$ gate) on the $j$th qubit~\cite{PhysRevA.93.050301,Puri2017}.
To suppress the unwanted $R_X$ rotations, we impose
\begin{linenomath}
\begin{align}
    \Delta_j(t)=\frac{g_{j\mathrm{c}}^2}{\Delta_{\mathrm{c}}(t)}
    \label{eq:noRX1}
\end{align}
\end{linenomath}
for all times $t$.

We focus on a subspace spanned by the following four degenerate eigenstates
of $\hat H_{0\mathrm{th}}(t)$:
\begin{linenomath}
\begin{align}
    \ket{\psi_{0,0}(t)}
    &=
    \ket{\alpha_1,\alpha_2,-\alpha_{\mathrm c}^{+}(t)},
    \label{eq:psi00t}\\
    \ket{\psi_{0,1}(t)}
    &=
    \ket{\alpha_1,-\alpha_2,-\alpha_{\mathrm c}^{-}(t)},
    \\
    \ket{\psi_{1,0}(t)}
    &=
    \ket{-\alpha_1,\alpha_2,\alpha_{\mathrm c}^{-}(t)},
    \\
    \ket{\psi_{1,1}(t)}
    &=
    \ket{-\alpha_1,-\alpha_2,\alpha_{\mathrm c}^{+}(t)},
    \label{eq:psi11t}
\end{align}
\end{linenomath}
where
\begin{linenomath}
\begin{align}
    \alpha_{\mathrm c}^{\pm}(t)
    =
    \frac{
        g_{1\mathrm c}\alpha_1
        \pm
        g_{2\mathrm c}\alpha_2
    }
    {\Delta_{\mathrm c}(t)},
    \label{eq:alphacpmt1}
\end{align}
\end{linenomath}
and $\alpha_j=\sqrt{p_j/K_j}$.
These four states share the same eigenenergy $E^{(0)}:=\sum_{j=1,2}\hbar K_j\alpha_j^4/2$.
We define the computational basis of the two Kerr-cat qubits by
$\{|\widetilde{k,l}\rangle:=|\psi_{k,l}(0)\rangle
\mid k,l=0,1\}$.

We use $\hat{H}_{ZZ}(t)$ to implement an
$R_{ZZ}$ gate while suppressing unwanted interactions
at the beginning and end of the protocol.
To this end, we set
$\Delta_{\mathrm{c}}(0)=\Delta_{\mathrm{c}}(t_{\mathrm{g}})=g_{1\mathrm{c}}g_{2\mathrm{c}}/g_{12}$,
which ensures
$\hat{H}_{ZZ}(0)=\hat{H}_{ZZ}(t_{\mathrm{g}})=0$,
where $t_{\mathrm{g}}$ is the gate time.

The $R_{ZZ}$ gate is implemented by varying the coupler detuning
$\Delta_{\mathrm c}(t)$ in time.
When the variation is sufficiently slow,
the state approximately remains in the subspace spanned by
$\{
\ket{\psi_{0,0}(t)},
\ket{\psi_{0,1}(t)},
\ket{\psi_{1,0}(t)},
\ket{\psi_{1,1}(t)}
\}$.

In this study, we assume
$g_{1\mathrm{c}}\alpha_1=g_{2\mathrm{c}}\alpha_2$,
which yields
$\alpha_{\mathrm c}^{-}(t)=0$
for all $t$.
Hence, only $\ket{\psi_{0,0}(t)}$ and $\ket{\psi_{1,1}(t)}$ involve displacement of the coupler.

As shown in Appendix~\ref{app:RZZ}, when the evolution is adiabatic,
an $R_{ZZ}$ gate is realized with rotation angle
\begin{linenomath}
\begin{align}
    \Theta_{\mathrm{ad}}
    =
    2
    \int_0^{t_{\mathrm g}}
    \frac{E^{(1)}(t)}{\hbar}
    \,dt ,
    \label{eq:Thetaad1}
\end{align}
\end{linenomath}
where $t_{\mathrm g}$ is the gate time and
\begin{linenomath}
\begin{align}
    E^{(1)}(t):=2\hbar\alpha_1\alpha_2\left(g_{12}-\frac{g_{1\mathrm{c}}g_{2\mathrm{c}}}{\Delta_{\mathrm{c}}(t)}\right).
\end{align}
\end{linenomath}

The gate speed is limited by the requirement that the coupler follow the
instantaneous coherent state adiabatically.
Therefore, the problem is reduced to suppressing the nonadiabatic excitation associated with the coupler displacement in $\ket{\psi_{0,0}(t)}$ and $\ket{\psi_{1,1}(t)}$.
In Sec.~\ref{sec:Application}, we develop a protocol for this purpose.

\subsection{Effective model of the coupler}
To accelerate the $R_{ZZ}$ gate while suppressing
nonadiabatic transitions, we focus on the dynamics
of the coupler.

When the KPO states remain close to the coherent
states appearing in Eqs.~\eqref{eq:psi00t}--\eqref{eq:psi11t},
the coupler dynamics can be described by effective
Hamiltonians obtained by taking expectation values
of the total Hamiltonian with respect to the KPO
degrees of freedom.
Namely, we define
\begin{linenomath}
\begin{align}
    \hat{H}_{\mathrm{c},0,0}^{\mathrm{eff}}(t)
    &:=\braket{\alpha_1,\alpha_2|\hat{H}(t)|\alpha_1,\alpha_2}
    \notag \\
    &=\hbar\Delta_{\mathrm{c}}(t)[\hat{a}_{\mathrm{c}}^{\dag}+\alpha_{\mathrm{c}}^{+}(t)][\hat{a}_{\mathrm{c}}+\alpha_{\mathrm{c}}^{+}(t)]+E_{0,0}(t), \label{eq:Hceff00} \\
    \hat{H}_{\mathrm{c},0,1}^{\mathrm{eff}}(t) 
    &:=\braket{\alpha_1,-\alpha_2|\hat{H}(t)|\alpha_1,-\alpha_2}
    \notag \label{eq:Hceff01}\\
    &=\hbar\Delta_{\mathrm{c}}(t)\hat{a}_{\mathrm{c}}^{\dag}\hat{a}_{\mathrm{c}}+E_{0,1}(t), \\
    \hat{H}_{\mathrm{c},1,0}^{\mathrm{eff}}(t) 
    &:=\braket{-\alpha_1,\alpha_2|\hat{H}(t)|-\alpha_1,\alpha_2}
    \notag \label{eq:Hceff10}\\
    &=\hbar\Delta_{\mathrm{c}}(t)\hat{a}_{\mathrm{c}}^{\dag}\hat{a}_{\mathrm{c}}+E_{1,0}(t), \\
    \hat{H}_{\mathrm{c},1,1}^{\mathrm{eff}}(t)
    &:=\braket{-\alpha_1,-\alpha_2|\hat{H}(t)|-\alpha_1,-\alpha_2}
    \notag \\
    &=\hbar\Delta_{\mathrm{c}}(t)[\hat{a}_{\mathrm{c}}^{\dag}-\alpha_{\mathrm{c}}^{+}(t)][\hat{a}_{\mathrm{c}}-\alpha_{\mathrm{c}}^{+}(t)]+E_{1,1}(t). \label{eq:Hceff11}
\end{align}
\end{linenomath}
Here, energy $E_{k,l}$ is defined by
\begin{linenomath}
\begin{align}
    \{E_{k,l}(t):=\langle\psi_{k,l}(t)|\hat{H}(t)|\psi_{k,l}(t)\rangle|k,l=0,1\},
\end{align}
\end{linenomath}
and contributes only to a state-dependent phase and does not affect
the coupler displacement [see also Eqs.~(\ref{eq:E00E11}) and (\ref{eq:E01E10})].

The effective Hamiltonians in Eqs.~(\ref{eq:Hceff00})--(\ref{eq:Hceff11})
have the same form as the Hamiltonian of an
off-resonantly driven resonator.
For example, $\hat H_{\mathrm c,1,1}^{\mathrm{eff}}(t)$ can be rewritten as
\begin{align}
\hat H_{\mathrm c,1,1}^{\mathrm{eff}}(t)
&=
\hbar\Delta_{\mathrm c}(t)
\hat a_{\mathrm c}^{\dagger}
\hat a_{\mathrm c}
-
\hbar\Omega_{\mathrm c}
(
\hat a_{\mathrm c}^{\dagger}
+
\hat a_{\mathrm c}
)
\nonumber\\
&\quad
+
\hbar\Delta_{\mathrm c}(t)
\bigl[\alpha_{\mathrm c}^{+}(t)\bigr]^2
+
E_{1,1}(t),
\end{align}
where
\begin{align}
\Omega_{\mathrm c}
:=
\Delta_{\mathrm c}(t)\alpha_{\mathrm c}^{+}(t)
=
g_{1\mathrm c}\alpha_1
+
g_{2\mathrm c}\alpha_2 .
\end{align}
Here, $\Omega_{\mathrm c}$ is independent of time.
The third term is a classical-number term and only contributes to an overall phase. Therefore, $\hat H_{\mathrm c,1,1}^{\mathrm{eff}}(t)$ is equivalent to
\begin{align}
\hat H_{\mathrm c}^{\mathrm{eff}}(t)
=
\hbar\Delta_{\mathrm c}(t)
\hat a_{\mathrm c}^{\dagger}
\hat a_{\mathrm c}
-
\hbar\Omega_{\mathrm c}
(
\hat a_{\mathrm c}^{\dagger}
+
\hat a_{\mathrm c}
),
\label{eq:Hc_eff_intro}
\end{align}
which is precisely the Hamiltonian of an off-resonantly driven resonator with detuning $\Delta_{\mathrm c}(t)$ and real drive amplitude $\Omega_{\mathrm c}$.

Importantly, the effective drive amplitude
$\Omega_{\mathrm c}$ is fixed by the couplings and
the coherent-state amplitudes of the KPOs,
whereas the detuning $\Delta_{\mathrm c}(t)$ is
experimentally tunable through the coupler frequency.
Consequently, the problem of accelerating the
$R_{ZZ}$ gate reduces to the problem of realizing
fast displacement of an off-resonantly driven
resonator by modulating its detuning.
In the following sections, we develop a fast-forward
protocol for this effective model.


\section{
Fast-forwarding displacement of a harmonic oscillator}
\label{sec:Review}
Before constructing a fast-forward protocol for the
effective coupler model introduced in
Sec.~\ref{sec:System},
we briefly review the fast-forward protocol for the
displacement of a particle in a harmonic potential
developed in Ref.~\cite{doi:10.1098/rspa.2009.0446}.
The protocol reviewed here serves as the starting
point for the driven-resonator protocols developed
in Secs.~\ref{sec:Omega} and \ref{sec:Delta}.

The Hamiltonian of the system is represented as
\begin{linenomath}
\begin{align}
    \hat{H}_0(t)&=\frac{\hat{p}^2}{2m}+\frac{m\omega^2}{2}[\hat{x}-x_{0}(t)]^2\quad(0\leq t \leq T_{\mathrm{f}}),
    \label{eq:H0t1}
\end{align}
\end{linenomath}
where $\hat{x}$ and $\hat{p}$ denote the position and momentum operators; $m$ is the mass; $\omega$ is the resonance frequency; $x_0$ characterizes the displacement of the potential; $T_{\mathrm{f}}$ is the final time of the displacement.  
The $n$th eigenenergy is $E_n=n\hbar\omega$ for all time.
Let $|\phi_n(t)\rangle$ denotes the $n$th energy eigenstate at time $t$.
If $x_0$ is varied sufficiently slowly, the adiabatic dynamics is realized.
The initial state is represented as
\begin{linenomath}
\begin{align}
    |\psi_{\rm ad}(0)\rangle=\sum_n c_n|\phi_n(0)\rangle,
    \label{eq:Psi_ad_ini1}
\end{align}
\end{linenomath}
where $\{c_n\}$ are coefficients.
The intermediate state in the adiabatic dynamics is given by
\begin{linenomath}
\begin{align}
    |\psi_{\rm ad}(t)\rangle = \sum_n c_n e^{-\mathrm{i}E_n  t/\hbar}|\phi_n(t)\rangle\quad(0\leq t \leq T_{\mathrm{f}}).
    \label{eq:Psiadt1}
\end{align}
\end{linenomath}
However, when the dynamics is nonadiabatic, the intermediate state $\Ket{\psi_{\mathrm{nonad}}(t)}$ differs from $|\psi_{\rm ad}(t)\rangle$. In general, the final state $\Ket{\psi_{\mathrm{nonad}}(T_{\mathrm{f}})}$ also differs from $\Ket{\psi_{\mathrm{ad}}(T_{\mathrm{f}})}$.

To obtain the final state of adiabatic dynamics under a general potential in a short time, Masuda and Nakamura devised the FFST~\cite{doi:10.1098/rspa.2009.0446}.
When a potential is harmonic, a fast-forward Hamiltonian for perfect displacement is represented as~\cite[Eq.~(3.11)]{doi:10.1098/rspa.2009.0446}
\begin{linenomath}
\begin{align}
    \hat{H}_{\mathrm{FF}}(t)&=\hat{H}_0(t)-m\ddot{x}_0(t)\hat{x},\nonumber\\
    &=\frac{\hat{p}^2}{2m}+\frac{m\omega^2}{2}[\hat{x}-x_{\rm FF}(t)]^2
    \label{eq:Hamiltonian_FF1}
\end{align}
\end{linenomath}
with
\begin{linenomath}
\begin{align}
x_{\rm FF}(t) = x_0(t) + \ddot{x}_0(t)/\omega^2.
\label{xFF_4_9_25}
\end{align}
\end{linenomath}
In the last equality in Eq.~\eqref{eq:Hamiltonian_FF1}, we ignored the difference of a classical-number term.
We do so except in Sec.~\ref{sec:Application} and Appendix~\ref{app:derivation}.
$\varepsilon\Lambda(t)$ and $\varepsilon\dot{\alpha}(t)$ in Ref.~\cite{doi:10.1098/rspa.2009.0446} correspond to $x_0(t)$ and $\ddot{x}_0(t)$ in this paper respectively.
Combining Eqs.~(2.24), (2.26), and (3.8) in Ref.~\cite{doi:10.1098/rspa.2009.0446}, we comprehend that the Hamiltonian $\hat{H}_{\rm FF}$ realizes the dynamics of the system represented as
\begin{linenomath}
\begin{align}
|\psi_{\rm FF}(t)\rangle = e^{\mathrm{i}f(\hat{x},t)} |\psi_{\rm ad}(t)\rangle, 
\label{PsiFF_4_9_25}
\end{align}
\end{linenomath}
where
\begin{linenomath}
\begin{align}
    f(\hat{x},t) = \frac{m}{\hbar}\dot{x}_0(t)\hat{x}.
\end{align}
\end{linenomath}
$\varepsilon\alpha(t)$ in Ref.~\cite{doi:10.1098/rspa.2009.0446} corresponds to $\dot{x}_0(t)$ in this paper. $\varepsilon$ is infinitesimal and $\alpha(t)$ is infinitely large in the reference.
In Eq.~(\ref{PsiFF_4_9_25}), we omitted space-independent, time-dependent overall phase, which is not relevant to the dynamics of the system.
Hereafter, we omit overall phases of other states without notice except in Sec.~\ref{sec:Application} and Appendix~\ref{app:derivation}.
By imposing $\dot{x}_0(0) = \dot{x}_0(T_{\mathrm{f}})=0$ in Eq.~\eqref{PsiFF_4_9_25}, we have $|\psi_{\rm FF}(0)\rangle=|\psi_{\rm ad}(0)\rangle$ and $|\psi_{\rm FF}(T_{\mathrm{f}})\rangle=|\psi_{\rm ad}(T_{\mathrm{f}})\rangle$.
In this way, we can generate $|\psi_{\rm ad}(T_{\mathrm{f}})\rangle$ for small $T_{\rm f}$.
Moreover, by imposing $\ddot{x}_0(0) = \ddot{x}_0(T_{\mathrm{f}})=0$, $\hat{H}_{\rm FF}$ coincides with $\hat{H}_0$ at the initial and final time.

\section{Perfect displacement via modulating drive amplitude}
\label{sec:Omega}
Sec.~\ref{sec:Review} can be reformulated in the language of driven resonators.
This reformulation serves as an intermediate step toward the effective coupler model discussed in
Sec.~\ref{sec:System}. We therefore consider an off-resonantly driven resonator with a time-dependent drive amplitude.

\begin{figure}
    \centering
    \includegraphics[width=0.31\textwidth]{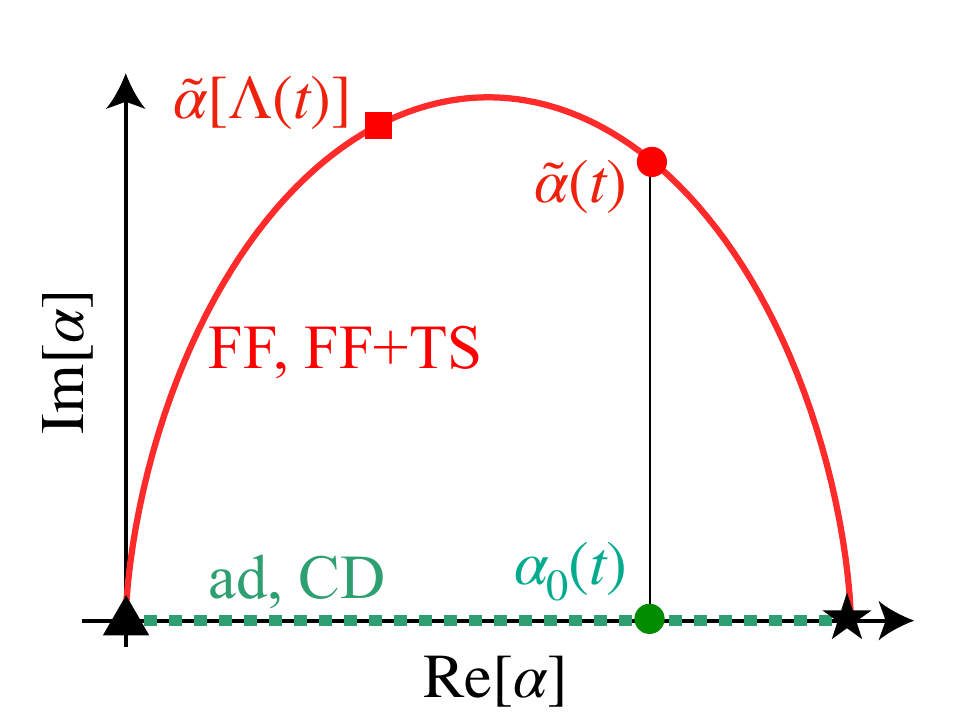}
    \caption{Schematic illustration of displacement trajectories under different types of control. The triangle and star represent the initial and final displacements, $\alpha_{\mathrm{i}}$ and $\alpha_{\mathrm{f}}$, respectively. The red solid curve and the green dashed line correspond to the trajectories of the fast-forwarded dynamics and the adiabatic dynamics, respectively. The latter coincides with the trajectory obtained using the counterdiabatic (CD) protocol. The dynamics based solely on the fast-forward (FF) protocol and that based on both the FF and time-scaling (TS) protocols follow the same trajectory at different speeds.
    The FF dynamics and the CD dynamics share the same velocity component in the $\mathrm{Re}[\alpha]$ direction. Note that the displacement under the FF dynamics, $\tilde{\alpha}(t)=\alpha_0(t)+\mathrm{i}\dot{\alpha}_0(t)$, and that under the CD dynamics, $\alpha_0(t)$, have the same argument $t$; $\tilde{\alpha}(0)=\alpha_0(0)=\alpha_{\mathrm{i}}$ and $\tilde{\alpha}(T_{\mathrm{f}})=\alpha_0(T_{\mathrm{f}})=\alpha_{\mathrm{f}}$. In contrast, the displacement under the $\mathrm{FF}+\mathrm{TS}$ dynamics, $\tilde{\alpha}[\Lambda(t)]$, have the scaled time, $\Lambda(t)$, which satisfies $\Lambda(0)=0$ and $\Lambda(t_{\mathrm{f}})=T_{\mathrm{f}}$; $\alpha_{\mathrm{f}}$ is reached at $t=t_{\mathrm{f}}$.
    \label{fig:trajectories}}
\end{figure}

Replacing $\omega$ in Hamiltonians \eqref{eq:H0t1} and \eqref{eq:Hamiltonian_FF1} with detuning $\Delta$ and using creation and annihilation operators,
\begin{linenomath}
\begin{align}
    \hat{a}^{\dag}&=\sqrt{\frac{m\Delta}{2\hbar}}\hat{x}-\frac{\mathrm{i}}{\sqrt{2m\hbar\Delta}}\hat{p}, \\
    \hat{a}&=\sqrt{\frac{m\Delta}{2\hbar}}\hat{x}+\frac{\mathrm{i}}{\sqrt{2m\hbar\Delta}}\hat{p},
\end{align}
\end{linenomath}
where $\hbar$ is the reduced Planck constant, we can transform the Hamiltonians into forms with tunable drive amplitude $\Omega$:
\begin{linenomath}
\begin{align}
    \hat{H}_0(t)/\hbar &= \Delta \hat{a}^\dagger \hat{a} - \Omega_0(t) (\hat{a}^\dagger + \hat{a})
    \notag \\
    &=\Delta\hat{D}[\alpha_0(t)]\hat{a}^\dagger \hat{a}\hat{D}^{\dag}[\alpha_0(t)]\quad(0\leq t \leq T_{\mathrm{f}}), \label{eq:H0t2} \\
    \hat{H}_{\rm FF}(t)/\hbar &= \Delta \hat{a}^\dagger \hat{a} - \Omega_{\rm FF}(t) (\hat{a}^\dagger + \hat{a}) \notag \\
    &=\Delta\hat{D}[\alpha_{\rm FF}(t)]\hat{a}^\dagger \hat{a}\hat{D}^{\dag}[\alpha_{\rm FF}(t)]\quad(0\leq t \leq T_{\mathrm{f}}),
    \label{HFF_drive_6_10_25}
\end{align}
\end{linenomath}
where
\begin{linenomath}
\begin{align}
    \Omega_0(t)&=\sqrt{\frac{m\Delta^3}{2\hbar}}x_0(t), \label{Omega_6_10_25} \\
    \alpha_0(t)&=\Omega_0(t)/\Delta, \\
    \Omega_{\rm FF}(t) &= \sqrt{\frac{m\Delta^3}{2\hbar}}x_{\rm FF}(t)=\Omega_0(t) + \ddot{\Omega}_0(t)/\Delta^2,
    \label{eq:OmegaFF1} \\
    \alpha_{\rm FF}(t)&=\Omega_{\rm FF}(t)/\Delta=\alpha_0(t)+\ddot{\alpha}_0(t)/\Delta^2,
\end{align}
\end{linenomath}
and $\hat{D}(\alpha)=\exp[\alpha\hat{a}^\dagger-\alpha^{*}\hat{a}]$ is a displacement operator.
We consider $\Omega_0(t)$ to be real, so that $\alpha_0(t)$, $\Omega_{\rm FF}(t)$, and $\alpha_{\rm FF}(t)$ are also real.
The boundary conditions in the previous section, $\dot{x}_0(0)=\dot{x}_0(T_{\mathrm{f}})=0$ and $\ddot{x}_0(0)=\ddot{x}_0(T_{\mathrm{f}})=0$, correspond to
\begin{linenomath}
\begin{gather}
    \dot{\Omega}_0(0)=\dot{\Omega}_0(T_{\mathrm{f}})=0, \quad
    \ddot{\Omega}_0(0)=\ddot{\Omega}_0(T_{\mathrm{f}})=0,
    \label{eq:Omega_boundary1} \\
    \dot{\alpha}_0(0)=\dot{\alpha}_0(T_{\mathrm{f}})=0, \quad
    \ddot{\alpha}_0(0)=\ddot{\alpha}_0(T_{\mathrm{f}})=0.
    \label{eq:alpha_boundary1}
\end{gather}
\end{linenomath}
From Eq.~\eqref{eq:H0t2}, we learn
\begin{linenomath}
\begin{align}
    |\phi_n(t)\rangle=\hat{D}[\alpha_0(t)]|n\rangle,\quad
    E_n=n\hbar\Delta,
\end{align}
\end{linenomath}
where $\ket{n}$ is a Fock state satisfying $\hat{a}^{\dag}\hat{a}\ket{n}=n\ket{n}$.
The adiabatic state in Eq.~\eqref{eq:Psiadt1} and the fast-forwarded state in Eq.~(\ref{PsiFF_4_9_25}) can be rewritten as
\begin{linenomath}
\begin{align}
    |\psi_{\rm ad}(t)\rangle&=\hat{D}[\alpha_0(t)]\sum_n c_n e^{-\mathrm{i}n \Delta t}|n\rangle, \label{eq:Psiadt2} \\
    |\psi_{\rm FF}(t)\rangle &= \hat{D}[\mathrm{i}\dot{\alpha}_0(t)/\Delta]|\psi_{\rm ad}(t)\rangle \notag \\
    &=\hat{D}[\tilde{\alpha}(t)]\sum_n c_n e^{-\mathrm{i}n \Delta t}|n\rangle,
    \label{eq:PsiFFt2}
\end{align}
\end{linenomath}
where $\tilde{\alpha}(t)$ is the displacement of the fast-forwarded state, represented as
\begin{linenomath}
\begin{align}
    \tilde{\alpha}(t)=\alpha_0(t)+\mathrm{i}\dot{\alpha}_0(t)/\Delta.
    \label{eq:tilde_alpha1}
\end{align}
\end{linenomath}
A different derivation of $|\psi_{\rm FF}(t)\rangle$ is given in Appendix~\ref{app:derivation}.
The difference of $\tilde{\alpha}(t)-\alpha_0(t)=\mathrm{i}\dot{\alpha}_0(t)/\Delta$ between the displacements of the above two states is imaginary for real $\alpha_0(t)$, as illustrated in Fig.~\ref{fig:trajectories}.

\begin{figure*}
    \centering
    \includegraphics[width=0.8\textwidth]{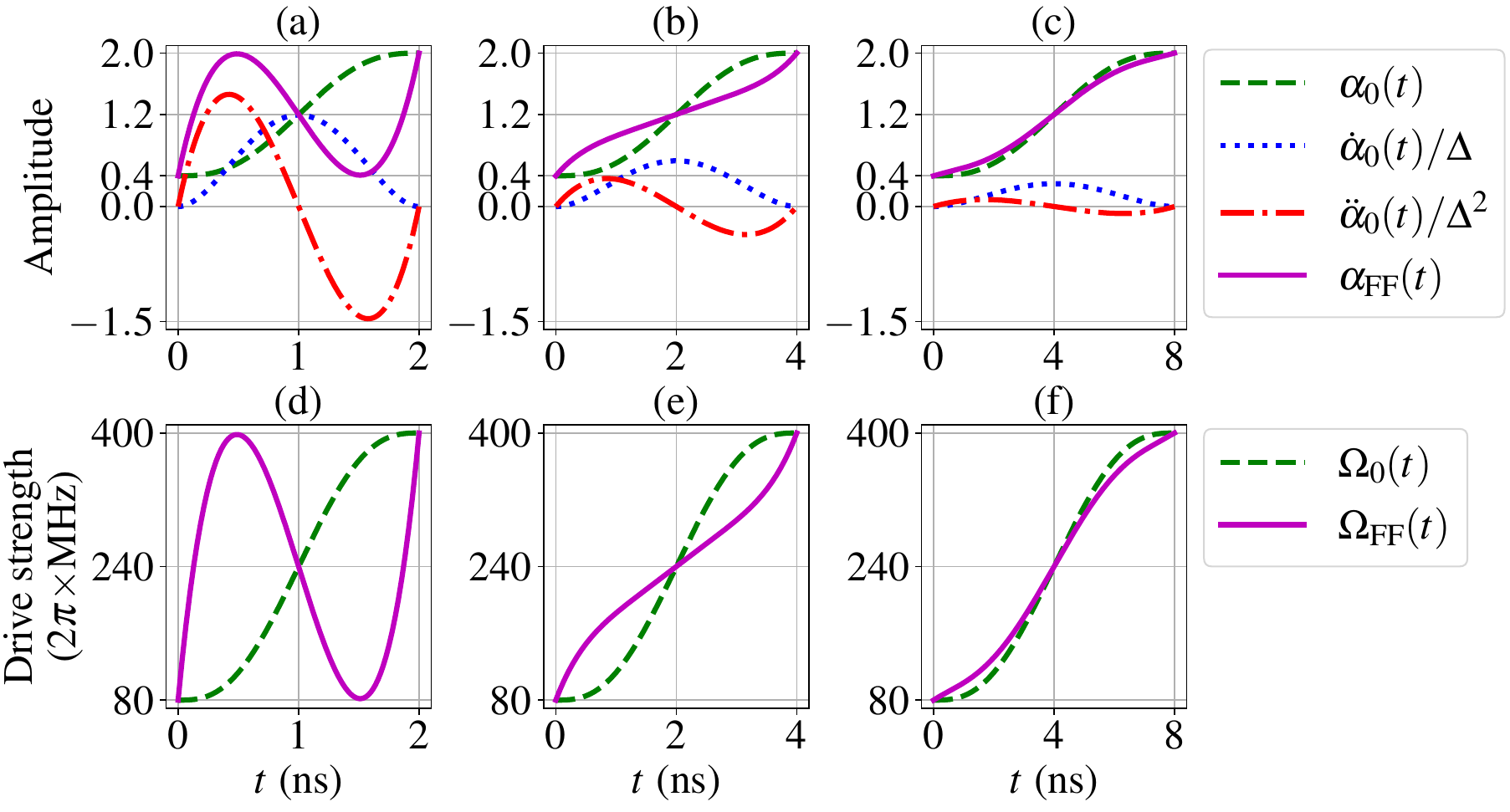}
    \caption{The time dependence of $\alpha_0(t)$, $\dot{\alpha}_0(t)/\Delta$, $\ddot{\alpha}_0(t)/\Delta^2$, and $\alpha_{\mathrm{FF}}(t)$ (a–c) and of $\Omega_0(t)$ and $\Omega_{\mathrm{FF}}(t)$ (d–f).
    The form of $\Omega_0(t)$ is given in Eq.~\eqref{eq:Omegat}.
    We set $\Delta/2\pi=200$ MHz, $\Omega_{\mathrm{i}}/2\pi=80$ MHz, and $\Omega_{\mathrm{f}}/2\pi=400$ MHz. $T_{\mathrm{f}}=2$ ns in (a) and (d), $T_{\mathrm{f}}=4$ ns in (b) and (e), and $T_{\mathrm{f}}=8$ ns in (c) and (f). 
    \label{fig:alphat1}}
\end{figure*}

To satisfy the boundary conditions of $\Omega_0(t)$ in Eqs.~\eqref{eq:Omega_boundary1}, we employ the following function~\cite{PhysRevA.83.013415}:
\begin{linenomath}
\begin{align}
    \Omega_0(t)&=\Omega_{\mathrm{i}}+(\Omega_{\mathrm{f}}-\Omega_{\mathrm{i}})g(t)\quad(0\leq t\leq T_{\mathrm{f}})
    \label{eq:Omegat}
\end{align}
\end{linenomath}
with
\begin{linenomath}
\begin{align}
    g(t):=10\left(\frac{t}{T_{\mathrm{f}}}\right)^3-15\left(\frac{t}{T_{\mathrm{f}}}\right)^4+6\left(\frac{t}{T_{\mathrm{f}}}\right)^5.
\end{align}
\end{linenomath}
We show the time dependence of $\alpha_0(t)$, $\dot{\alpha}_0(t)/\Delta$, $\ddot{\alpha}_0(t)/\Delta^2$, $\alpha_{\rm FF}(t)$, $\Omega_0(t)$, and $\Omega_{\mathrm{FF}}(t)$ in Fig.~\ref{fig:alphat1}.
As $T_{\mathrm{f}}$ decreases, the difference between $\alpha_0(t)$ and $\alpha_{\mathrm{FF}}(t)$ and that between $\Omega_0(t)$ and $\Omega_{\mathrm{FF}}(t)$ become more pronounced.

\begin{figure}
    \centering
    \includegraphics[width=0.36\textwidth]{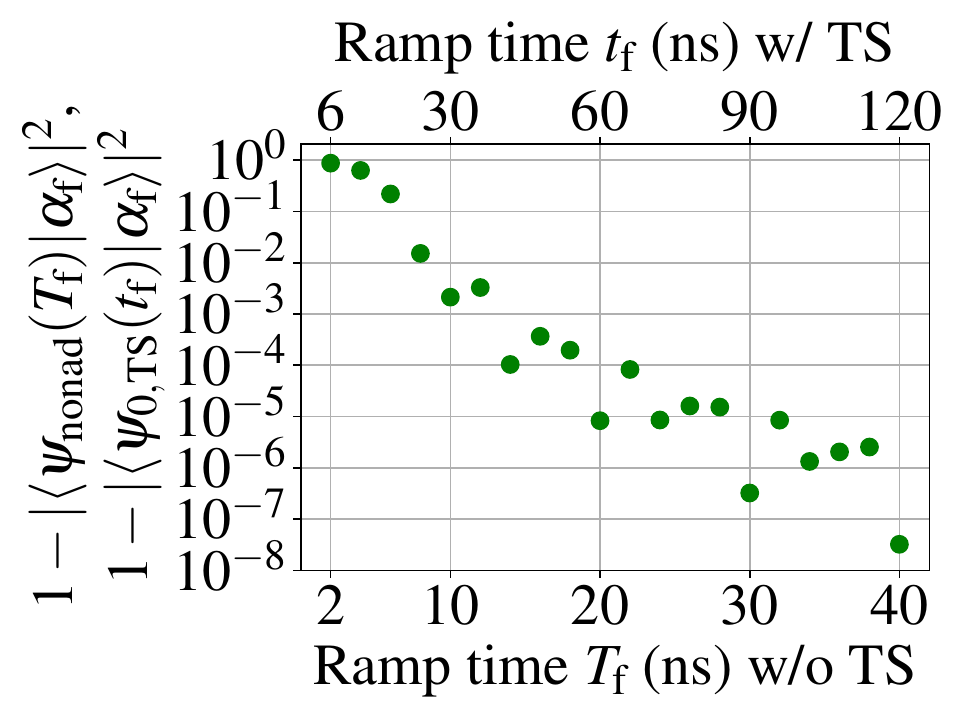}
    \caption{The infidelity between the target state $\ket{\alpha_{\mathrm{f}}}$ and the final state $\Ket{\psi_{\mathrm{nonad}}(T_{\mathrm{f}})}$ under $\hat{H}_0(t)$ in Eq.~\eqref{eq:H0t2} with $\Omega_0(t)$ in Eq.~\eqref{eq:Omegat}, $1-|\braket{\psi_{\mathrm{nonad}}(T_{\mathrm{f}})|\alpha_{\mathrm{f}}}|^2$,
     when $\Delta/2\pi=200$ MHz, $\Omega_{\mathrm{i}}/2\pi=80$ MHz, and $\Omega_{\mathrm{f}}/2\pi=400$ MHz; the infidelity between the target state $|\alpha_{\mathrm{f}}\rangle$ and the final state $|\psi_{0,\rm TS}(t_{\mathrm{f}})\rangle$ under $\hat{H}_{0,\rm TS}(t)$ in Eq.~\eqref{eq:H_0TS}, $1-|\braket{\psi_{0,\mathrm{TS}}(t_{\mathrm{f}})|\alpha_{\mathrm{f}}}|^2$, when  $\Delta_{\mathrm{i}}/2\pi=200$ MHz, $\Delta_{\mathrm{f}}/2\pi=40$ MHz, and $\Omega_{\mathrm{i}}/2\pi=80$ MHz.
    The latter infidelity is a time-scaled version of the former with the scaling in Eq.~\eqref{eq:tfTf1}, $t_{\mathrm{f}}=3T_{\mathrm{f}}$.
    Vanishing infidelity can be achieved using FF [Eq.~\eqref{HFF_drive_6_10_25}] or CD [eq.~\eqref{eq:HCD1}] for any ramp time $T_{\mathrm{f}}$ or using FF+TS [Eq.~\eqref{H_FFS_4_14_25}] for any ramp time $t_{\mathrm{f}}$ that satisfies $\Omega_{\rm FF}[\Lambda(t)]\neq0$ for $0\leq t\leq t_{\mathrm{f}}$.
    \label{fig:drive_infidelity1}}
\end{figure}

While we can obtain the final adiabatic state $|\psi_{\rm ad}(T_{\mathrm{f}})\rangle$ with unit fidelity even for small $T_{\mathrm{f}}$ under $\hat{H}_{\mathrm{FF}}(t)$ in Eq.~\eqref{HFF_drive_6_10_25}, we cannot under $\hat{H}_0(t)$ in Eq.~\eqref{eq:H0t2}.
In the case that the initial state is $\Ket{\psi_{\mathrm{nonad}}(0)}=\Ket{\alpha_0(0)}=\Ket{\alpha_{\mathrm{i}}}$, which is identical to $\Ket{\psi_{\mathrm{ad}}(0)}$ in Eq.~\eqref{eq:Psiadt2} with $c_n=\delta_{n,0}$, the infidelity between the final state $\Ket{\psi_{\mathrm{nonad}}(T_{\mathrm{f}})}$ and the target state $\Ket{\alpha_0(T_{\mathrm{f}})}=\Ket{\alpha_{\mathrm{f}}}$, $1-|\braket{\psi_{\mathrm{nonad}}(T_{\mathrm{f}})|\alpha_{\mathrm{f}}}|^2$, is numerically calculated as in Fig.~\ref{fig:drive_infidelity1}.
As $T_{\mathrm{f}}$ decreases, the infidelity tends to increase.
Numerical calculations in this paper were performed using Quantum Toolbox in Python (QuTiP)~\cite{QuTiP2012,QuTiP2013,LAMBERT20261}

As shown in Appendix~\ref{sec:CD}, the counterdiabatic (CD) Hamiltonian~\cite{PhysRevLett.85.1626,Demirplak2003,Demirplak2005,Berry_2009,doi:10.1098/rsta.2021.0272} represented as
\begin{linenomath}
\begin{align}
    \hat{H}_{\rm CD}(t)/\hbar&=\Delta\hat{D}[\alpha_{\mathrm{CD}}(t)]\hat{a}^{\dag}\hat{a}\hat{D}^{\dag}[\alpha_{\mathrm{CD}}(t)] \notag \\ 
    &=\Delta\hat{a}^{\dag}\hat{a}-[\Omega_{\mathrm{CD}}(t)\hat{a}^{\dag}+\Omega_{\mathrm{CD}}^{*}(t)\hat{a}],
    \label{eq:HCD1}
\end{align}
\end{linenomath}
where
\begin{linenomath}
\begin{align}
    \alpha_{\mathrm{CD}}(t)&=\alpha_0(t)-\mathrm{i}\dot{\alpha}_0(t)/\Delta, \\
    \Omega_{\mathrm{CD}}(t)&=\Delta\alpha_{\mathrm{CD}}(t)=\Omega_0(t)-\mathrm{i}\dot{\Omega}_0(t)/\Delta,
    \label{eq:OmegaCD1}
\end{align}
\end{linenomath}
generates $|\psi_{\rm ad}(t)\rangle$.
Thus, the CD method also yields the final adiabatic state $|\psi_{\mathrm{ad}}(T_{\mathrm{f}})\rangle$ with unit fidelity even for small $T_{\mathrm{f}}$.

\section{Perfect displacement via modulating detuning}
\label{sec:Delta}
We now develop a protocol that achieves perfect
displacement by controlling the detuning $\Delta$
rather than the drive amplitude $\Omega$.
Although resonant driving can realize perfect displacement of a resonator, it is incompatible with the coupler dynamics required for the tunable-coupler-based $R_{ZZ}$ gate considered here (see Appendices~\ref{app:resonant} and~\ref{sec:resonant}).
This situation is relevant to the
effective coupler model introduced in
Sec.~\ref{sec:System}, where the experimentally
accessible control parameter is $\Delta$.
Our approach combines the fast-forward-scaling and time-scaling approaches to construct a detuning
trajectory that realizes perfect displacement.
Readers interested primarily in the resulting
detuning profile may skip the derivation and proceed
directly to the paragraph containing
Eq.~\eqref{eq:ut2}.

\subsection{Time scaling}
We explain the concept of time scaling.
Let $|\psi(t)\rangle$ denote reference dynamics realized by a Hamiltonian $\hat{H}(t)$.
We consider time-scaled dynamics
\begin{linenomath}
\begin{align}
|\psi_{\rm TS}(t)\rangle = |\psi[\Lambda(t)]\rangle\quad(0\leq t\leq t_{\mathrm{f}}),
\end{align}
\end{linenomath}
where $\Lambda(t)$ is the scaled time defined as
\begin{linenomath}
\begin{align}
\Lambda(t)=\int_0^t S(s)\,\mathrm{d}s
\label{Lam_4_15_25}
\end{align}
\end{linenomath}
and satisfies $\Lambda(t_{\mathrm{f}})=T_{\mathrm{f}}$;
$S(t)$ is a time-dependent scaling factor.
Note that $\Lambda(0)=0$ by definition.
It is easily confirmed that we can realize the time-scaled dynamics by using the Hamiltonian defined as
\begin{linenomath}
\begin{align}
\hat{H}_{\rm TS}(t)=S(t)\hat{H}[\Lambda(t)].
\end{align}
\end{linenomath}

\subsection{Combination of fast-forward scaling and time scaling}
We apply the time scaling to the fast-forwarded dynamics, $|\psi_{\rm FF}(t)\rangle$ in Eq.~\eqref{eq:PsiFFt2}.
The time-scaled dynamics, $|\psi_{\rm FF,TS}(t)\rangle=|\psi_{\rm FF}[\Lambda(t)]\rangle$, is realized by $\hat{H}_{\mathrm{FF,TS}}(t)=S(t)\hat{H}_{\mathrm{FF}}[\Lambda(t)]$, where $\hat{H}_{\mathrm{FF}}$ is given in Eq.~\eqref{HFF_drive_6_10_25}.
To make the drive amplitude of $\hat{H}_{\mathrm{FF,TS}}(t)$ time independent, we set
\begin{linenomath}
\begin{align}
S(t) = \frac{\Omega_0(0)}{\Omega_{\rm FF}[\Lambda(t)]},
\label{S_4_14_25}
\end{align}
\end{linenomath}
which leads to
\begin{linenomath}
\begin{align}
&\quad\hat{H}_{\rm FF,TS}(t)/\hbar=\Delta_{\rm FF,TS}(t) \hat{a}^\dagger \hat{a} 
- \Omega_0(0) (\hat{a}^\dagger + \hat{a}) \notag \\
&=\Delta_{\rm FF,TS}(t)\hat{D}\{\alpha_{\rm FF}[\Lambda(t)]\}\hat{a}^\dagger \hat{a}\hat{D}^{\dag}\{\alpha_{\rm FF}[\Lambda(t)]\}
\label{H_FFS_4_14_25}
\end{align}
\end{linenomath}
for $0\leq t \leq t_{\mathrm{f}}$, where
\begin{linenomath}
\begin{align}
\Delta_{\rm FF,TS}(t) = \Delta \frac{\Omega_0(0)}{\Omega_{\rm FF}[\Lambda(t)]}.
\label{Delta_FFS_4_16_25}
\end{align}
\end{linenomath}
From Eqs.~\eqref{Lam_4_15_25} and \eqref{S_4_14_25}, $\Lambda(t)$ must satisfy
\begin{linenomath}
\begin{align}
\dot{\Lambda}(t)= \frac{\Omega_0(0)}{\Omega_{\rm FF}[\Lambda(t)]}.
\label{dLam_4_15_25}
\end{align}
\end{linenomath}

From Eqs.~\eqref{S_4_14_25} and \eqref{Delta_FFS_4_16_25}, we see that $\Omega_{\rm FF}[\Lambda(t)]$ must remain nonzero throughout the protocol.
This requirement does not arise when $\Omega$ itself is used as the control parameter.
Despite this additional requirement, as we will see in the next section, when the system of interest is a subsystem of a complex system, there are cases where modulating $\Delta$ rather than $\Omega$ is experimentally feasible.

Note that we can obtain $\hat{H}_{\rm FF,TS}(t)$ in a more general form where both the detuning and drive amplitude are time dependent by using $S(t)$ different from Eq.~(\ref{S_4_14_25}). However, modulating both is experimentally cumbersome and does not improve fidelity compared to modulating either, which can achieve perfect displacement.
We also note that combining the counterdiabatic and time-scaling approaches does not satisfy both a tunable real detuning and a fixed drive amplitude, because $\Omega_{\mathrm{CD}}(t)$ is complex.

Let us explain the procedure to design the time dependence of $\Delta_{\rm FF,TS}(t)$ in the case that $\Delta_{\rm FF,TS}(0)=\Delta_{\rm i}$, $\Delta_{\rm FF,TS}(t_{\mathrm{f}})=\Delta_{\rm f}$, and $\Omega_0(0)=\Omega_{\rm i}$.
First, we choose the time dependence of $\Omega_0(t)$ to satisfy Eqs.~\eqref{eq:Omega_boundary1}, $\Omega_0(0)=\Omega_{\rm i}$, and $\Omega_0(T_{\mathrm{f}})=\Omega_{\rm i}\Delta_{\rm i}/\Delta_{\rm f}$, which is obtained by substituting $t=t_{\mathrm{f}}$ into Eq.~\eqref{Delta_FFS_4_16_25}.
Then, we integrate Eq.~(\ref{dLam_4_15_25}) using Eq.~\eqref{eq:OmegaFF1} to obtain $\Lambda(t)$.
Finally, we obtain $\Delta_{\rm FF,TS}(t)$ from Eq.~\eqref{Delta_FFS_4_16_25}.
For example, when $\Omega_0(t)$ is chosen as in Eq.~\eqref{eq:Omegat} with $\Omega_{\mathrm{f}}=\Omega_{\rm i}\Delta_{\rm i}/\Delta_{\rm f}$, Eq.~(\ref{dLam_4_15_25}) is written as
\begin{linenomath}
\begin{align}
    \dot{\Lambda}(t)
    &=\left\{1+\left(\frac{\Delta_{\mathrm{i}}}{\Delta_{\mathrm{f}}}-1\right)
    \left[g[\Lambda(t)]+\frac{60w[\Lambda(t)]}{\Delta_{\mathrm{i}}^2T_{\mathrm{f}}^2}\right]\right\}^{-1}
    \label{eq:ut2}
\end{align}
\end{linenomath}
with
\begin{linenomath}
\begin{align}
    w(t):=\frac{t}{T_{\mathrm{f}}}-3\left(\frac{t}{T_{\mathrm{f}}}\right)^2+2\left(\frac{t}{T_{\mathrm{f}}}\right)^3.
\end{align}
\end{linenomath}
Integrating Eq.~\eqref{eq:ut2} over time yields
\begin{linenomath}
\begin{align}
    \Lambda(t)+\left(\frac{\Delta_{\mathrm{i}}}{\Delta_{\mathrm{f}}}-1\right)\left[G[\Lambda(t)]+\frac{60W[\Lambda(t)]}{\Delta_{\mathrm{i}}^2T_{\mathrm{f}}^2}\right]=t
    \label{eq:ut3}
\end{align}
\end{linenomath}
with
\begin{linenomath}
\begin{align}
    G(t)&:=\int_0^tg(s)\,\mathrm{d}s
    \notag \\
    &=T_{\mathrm{f}}\left[\frac{5}{2}\left(\frac{t}{T_{\mathrm{f}}}\right)^4-3\left(\frac{t}{T_{\mathrm{f}}}\right)^5+\left(\frac{t}{T_{\mathrm{f}}}\right)^6\right],
    \\
    W(t)&:=\int_0^tw(s)\,\mathrm{d}s
    \notag \\
    &=T_{\mathrm{f}}\left[\frac{1}{2}\left(\frac{t}{T_{\mathrm{f}}}\right)^2-\left(\frac{t}{T_{\mathrm{f}}}\right)^3+\frac{1}{2}\left(\frac{t}{T_{\mathrm{f}}}\right)^4\right].
\end{align}
\end{linenomath}
Substituting $t=t_{\mathrm{f}}$ into Eq.~\eqref{eq:ut3}, we obtain
\begin{linenomath}
\begin{align}
    t_{\mathrm{f}}=\frac{1}{2}\left(\frac{\Delta_{\mathrm{i}}}{\Delta_{\mathrm{f}}}+1\right)T_{\mathrm{f}}.
    \label{eq:tfTf1}
\end{align}
\end{linenomath}
By numerically solving Eq.~\eqref{eq:ut3}, we can obtain $\Lambda(t)$ for $0\leq t\leq t_{\mathrm{f}}$.

We can obtain the final adiabatic state $|\psi_{\rm ad}(T_{\mathrm{f}})\rangle=|\psi_{\rm FF,TS}(t_{\mathrm{f}})\rangle$ with unit fidelity even for small $t_{\mathrm{f}}$ under $\hat{H}_{\mathrm{FF,TS}}(t)$ in Eq.~\eqref{H_FFS_4_14_25}, as long as $\Omega_{\rm FF}[\Lambda(t)]\neq0$ for $0\leq t\leq t_{\mathrm{f}}$.
For comparison, we numerically calculate the infidelity between the target coherent state $|\alpha_{\mathrm{f}}\rangle$ and the final state of the time-scaled dynamics without FF, $|\psi_{0,\rm TS}(t_{\mathrm{f}})\rangle=|\psi_{\rm nonad}[\Lambda_0(t_{\mathrm{f}})]\rangle$, obtained under the Hamiltonian
\begin{linenomath}
\begin{align}
&\quad\hat{H}_{0,\rm TS}(t)/\hbar=\frac{\Omega_0(0)\hat{H}_0[\Lambda_0(t)]}{\hbar\Omega_{0}[\Lambda_0(t)]} \notag \\
&=\Delta_{0,\rm TS}(t) \hat{a}^\dagger \hat{a} 
- \Omega_0(0) (\hat{a}^\dagger + \hat{a}) \notag \\
&=\Delta_{0,\rm TS}(t)\hat{D}\{\alpha_0[\Lambda_0(t)]\}\hat{a}^\dagger \hat{a}\hat{D}^{\dag}\{\alpha_0[\Lambda_0(t)]\},
\label{eq:H_0TS}
\end{align}
\end{linenomath}
where
\begin{linenomath}
\begin{align}
\Delta_{0,\rm TS}(t) = \Delta \frac{\Omega_0(0)}{\Omega_{0}[\Lambda_0(t)]}
\label{eq:Delta_0TS}
\end{align}
\end{linenomath}
and $\Lambda_0(t)$ satisfies $\Lambda_0(0)=0$, $\Lambda_0(t_{\mathrm{f}})=T_{\mathrm{f}}$, and
\begin{linenomath}
\begin{align}
    \Lambda_0(t)+\left(\frac{\Delta_{\mathrm{i}}}{\Delta_{\mathrm{f}}}-1\right)G[\Lambda_0(t)]=t,
    \label{eq:Lambda0}
\end{align}
\end{linenomath}
starting from $|\psi_{0,\rm TS}(0)\rangle=|\alpha_{\mathrm{i}}\rangle$; see Fig.~\ref{fig:drive_infidelity1}.
Note that $\Delta_{0,\rm TS}(0)=\Delta_{\mathrm{i}}$, that $\Delta_{0,\rm TS}(t_{\mathrm{f}})=\Delta_{\mathrm{f}}$, and that substituting $t=t_{\mathrm{f}}$ into Eq.~\eqref{eq:Lambda0} also yields Eq.~\eqref{eq:tfTf1}.

\section{Fast-forwarded \texorpdfstring{$R_{ZZ}$}{RZZ} gate}
\label{sec:Application}
As discussed in Sec.~\ref{sec:System},
implementation of an $R_{ZZ}$ gate is reduced to controlling
the displacement of the coupler coherent state.
We now apply the detuning-control protocol developed
in Sec.~\ref{sec:Delta} to this effective model.

Under $\hat H_{\mathrm c,0,1}^{\mathrm{eff}}(t)$ and $\hat H_{\mathrm c,1,0}^{\mathrm{eff}}(t)$,
the state of the coupler remains in the vacuum state. Therefore, only $\hat H_{\mathrm c,0,0}^{\mathrm{eff}}(t)$ and $\hat H_{\mathrm c,1,1}^{\mathrm{eff}}(t)$ require control of the coupler displacement.
In what follows, we apply the fast-forward protocol to these two Hamiltonians.

\subsection{Fast and high-fidelity displacement of the coupler}
The effective Hamiltonian
$\hat{H}_{\mathrm c,1,1}^{\mathrm{eff}}(t)$
in Eq.~\eqref{eq:Hceff11}
has the same form as the driven-resonator Hamiltonian
studied in Sec.~\ref{sec:Delta}.
We therefore engineer the coupler detuning
$\Delta_{\mathrm c}(t)$ so that
$\hat{H}_{\mathrm c,1,1}^{\mathrm{eff}}(t)$
reproduces the fast-forward Hamiltonian
\eqref{H_FFS_4_14_25}
with the identification $\Omega_0(0)=g_{1,\mathrm{c}}\alpha_1+g_{2,\mathrm{c}}\alpha_2$.

Specifically, for
$0\le t\le t_{\mathrm g}/2$,
we choose
\begin{linenomath}
\begin{align}
    \begin{aligned}
        \Delta_{\mathrm{c}}(t)&=\Delta_{\mathrm{FF,TS}}(t) \\
        &=\Delta_{\mathrm{i}}\left\{1+\left(\frac{\Delta_{\mathrm{i}}}{\Delta_{\mathrm{f}}}-1\right)
        \left[g[\Lambda(t)]+\frac{60w[\Lambda(t)]}{\Delta_{\mathrm{i}}^2T_{\mathrm{f}}^2}\right]\right\}^{-1}
    \end{aligned} 
    \label{eq:DeltacFFSt1}
\end{align}
\end{linenomath}
with
\begin{linenomath}
\begin{align}
    \frac{t_{\mathrm{g}}}{2}=t_{\mathrm{f}}=\frac{1}{2}\left(\frac{\Delta_{\mathrm{i}}}{\Delta_{\mathrm{f}}}+1\right)T_{\mathrm{f}}.
\end{align}
\end{linenomath}
This $\Delta_{\mathrm{c}}(t)$ achieves perfect displacement not only from $\ket{\alpha_{\mathrm{c}}^{+}(0)}$ to $\ket{\alpha_{\mathrm{c}}^{+}(t_{\mathrm{g}}/2)}$ under $\hat{H}_{\mathrm{c},1,1}^{\mathrm{eff}}(t)$ but also from $\ket{-\alpha_{\mathrm{c}}^{+}(0)}$ to $\ket{-\alpha_{\mathrm{c}}^{+}(t_{\mathrm{g}}/2)}$ under $\hat{H}_{\mathrm{c},0,0}^{\mathrm{eff}}(t)$.
Analogously, by setting $\Delta_{\mathrm{c}}(t)=\Delta_{\mathrm{FF,TS}}(t_{\mathrm{g}}-t)$ for $t_{\mathrm{g}}/2\leq t\leq t_{\mathrm{g}}$, we can realize perfect displacement from $\ket{\pm\alpha_{\mathrm{c}}^{+}(t_{\mathrm{g}}/2)}$ to $\ket{\pm\alpha_{\mathrm{c}}^{+}(t_{\mathrm{g}})}=\ket{\pm\alpha_{\mathrm{c}}^{+}(0)}$.
The fast-forwarded state under $\hat{H}_{\mathrm{c},k,k}^{\mathrm{eff}}(t)$ ($k\in\{0,1\}$) is [see Eq.~\eqref{eq:psiFFt_gphase1}]
\begin{linenomath}
\begin{align}
    \ket{\psi_{\mathrm{c},k,k}^{\mathrm{eff}}(t)}&=\exp\left(-\frac{\mathrm{i}}{\hbar}\int_0^tE_{k,k}(s)\,\mathrm{d}s\right) \notag \\
    &\quad\times\exp\left(-\mathrm{i}\int_0^{\Lambda(t)}b(s)\,\mathrm{d}s\right)\ket{(-1)^{k+1}\tilde{\alpha}[\Lambda(t)]}
\end{align}
\end{linenomath}
for $0\leq t\leq t_{\mathrm{g}}/2$ and
\begin{linenomath}
\begin{align}
    \ket{\psi_{\mathrm{c},k,k}^{\mathrm{eff}}(t)}&=\exp\left(-\frac{\mathrm{i}}{\hbar}\int_0^tE_{k,k}(s)\,\mathrm{d}s\right) \notag \\
    &\quad\times\exp\left(-\mathrm{i}\int_0^{\Lambda(t_{\mathrm{g}}/2)}b(s)\,\mathrm{d}s\right) \notag \\
    &\quad\times
    \exp\left(-\mathrm{i}\int_{\Lambda(t_{\mathrm{g}}-t)}^{\Lambda(t_{\mathrm{g}}/2)}b(s)\,\mathrm{d}s\right)
    \notag \\
    &\quad\times
    \ket{(-1)^{k+1}\tilde{\alpha}[\Lambda(t_{\mathrm{g}}-t)]}
\end{align}
\end{linenomath}
for $t_{\mathrm{g}}/2\leq t\leq t_{\mathrm{g}}$, where
\begin{linenomath}
\begin{align}
    b(s)&=\alpha_{\mathrm{FF}}(s)\ddot{\alpha}_0(s)/\Delta_{\mathrm{i}}, \\
    \ket{\tilde{\alpha}[\Lambda(t)]}&=\ket{\alpha_0[\Lambda(t)]+\mathrm{i}\dot{\alpha}_0[\Lambda(t)]/\Delta_{\mathrm{i}}}, \label{eq:tilde_alpha_lambda1} \\
    \alpha_0[\Lambda(t)]&=\frac{g_{1,\mathrm{c}}\alpha_1+g_{2,\mathrm{c}}\alpha_2}{\Delta_{\mathrm{i}}}
    \left\{1+\left(\frac{\Delta_{\mathrm{i}}}{\Delta_{\mathrm{f}}}-1\right)g[\Lambda(t)]\right\}.
\end{align}
\end{linenomath}
In the case that the initial state of the system is given as in Eq.~\eqref{eq:app_initial1}, 
the intermediate state approximately lies in the subspace spanned by the following four states:
\begin{linenomath}
\begin{align}
    \ket{\psi^{\mathrm{FF}}_{0,0}(t)}&:=\ket{\alpha_1,\alpha_2,\psi_{\mathrm{c},0,0}^{\mathrm{eff}}(t)}, \label{eq:psiFF00t} \\
    \ket{\psi^{\mathrm{FF}}_{0,1}(t)}&:=\ket{\alpha_1,-\alpha_2,\psi_{\mathrm{c},0,1}^{\mathrm{eff}}(t)}, \\
    \ket{\psi^{\mathrm{FF}}_{1,0}(t)}&:=\ket{-\alpha_1,\alpha_2,\psi_{\mathrm{c},1,0}^{\mathrm{eff}}(t)}, \\
    \ket{\psi^{\mathrm{FF}}_{1,1}(t)}&:=\ket{-\alpha_1,-\alpha_2,\psi_{\mathrm{c},1,1}^{\mathrm{eff}}(t)}, \label{eq:psiFF11t}
\end{align}
\end{linenomath}
where
\begin{linenomath}
\begin{align}
    \ket{\psi_{\mathrm c,0,1}^{\mathrm{eff}}(t)}
    =
    \ket{\psi_{\mathrm c,1,0}^{\mathrm{eff}}(t)}
    =
    \exp\!\left(
        -\frac{\mathrm i}{\hbar}
        \int_0^t E_{0,1}(s)\,\mathrm ds
    \right)
    \ket{0}.
\end{align}
\end{linenomath}
The state at $t=t_{\mathrm{g}}$ is approximately written as
\begin{linenomath}
\begin{align}
    \ket{\Psi(t_{\mathrm{g}})}\approx\mathrm{e}^{-\mathrm{i}\theta_{\mathrm{FF}}}\hat{R}_{ZZ}(\Theta_{\mathrm{FF}})\ket{\Psi(0)}=:\ket{\Psi_{\mathrm{FF}}(t_{\mathrm{g}})},
\end{align}
\end{linenomath}
where
\begin{linenomath}
\begin{align}
    \theta_{\mathrm{FF}}&:=\frac{E^{(0)}t_{\mathrm{g}}}{\hbar}+\int_0^{\Lambda(t_{\mathrm{g}}/2)}b(t)\,\mathrm{d}t, \\
    \Theta_{\mathrm{FF}}&:=2\int_0^{t_{\mathrm{g}}}\frac{E^{(1)}(t)}{\hbar}\,\mathrm{d}t+2\int_0^{\Lambda(t_{\mathrm{g}}/2)}b(t)\,\mathrm{d}t. \label{eq:ThetaFF}
\end{align}
\end{linenomath}
Thus, the ${R}_{ZZ}(\Theta_{\mathrm{FF}})$ gate is implemented in a fast-forward manner.

\begin{figure}
    \centering
    \includegraphics[width=0.46\textwidth]{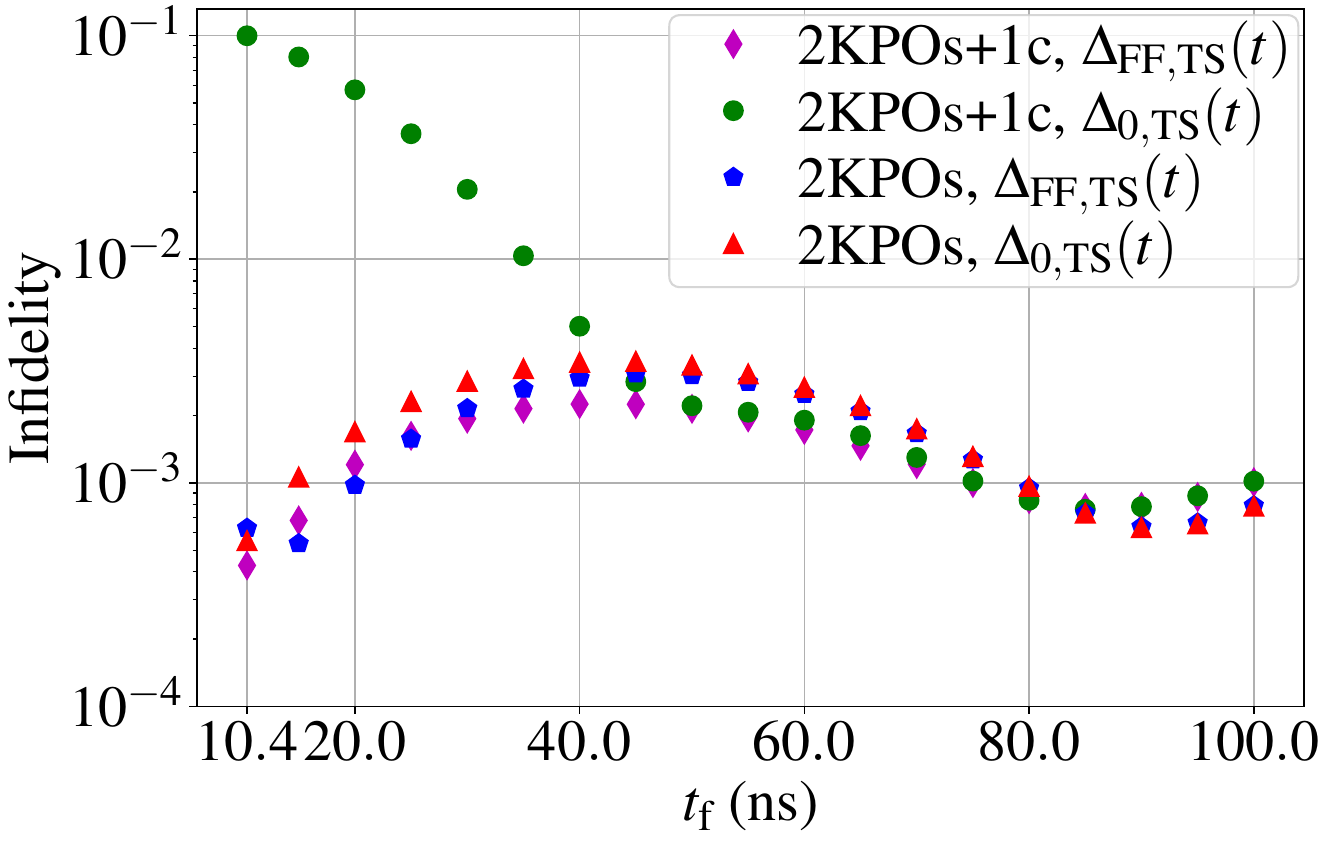}
    \caption{The infidelity $1-F(t_{\mathrm{f}})$ with $F(t_{\mathrm{f}})$ in Eq.~\eqref{eq:fidelity1} when $\Delta_{\mathrm{c}}(t)=\Delta_{\mathrm{FF,TS}}(t)$ in Eq.~\eqref{eq:DeltacFFSt1} (the magenta diamonds) and when $\Delta_{\mathrm{c}}(t)=\Delta_{0,\mathrm{TS}}(t)$ in Eq.~\eqref{eq:Deltac0TS2} (the green circles); the infidelity $1-\tilde{F}(t_{\mathrm{f}})$ with $\tilde{F}(t_{\mathrm{f}})$ in Eq.~\eqref{eq:fidelity2} when $\Delta_{\mathrm{c}}(t)=\Delta_{\mathrm{FF,TS}}(t)$ (the blue pentagons) and when $\Delta_{\mathrm{c}}(t)=\Delta_{0,\mathrm{TS}}(t)$ (the red triangles).
    The parameter values used are listed in Table~\ref{table:parameter1}.
    \label{fig:infidelity1}}
\end{figure}

To examine the effectiveness of our method, we perform numerical simulations.
We prepare the initial state of the system as $\ket{\Psi(0)}=\ket{\widetilde{1,1}}$.
The system time-evolves under $\hat{H}(t)$ in Eq.~\eqref{eq:Hamiltonian_totalR} until $t=t_{\mathrm{f}}$ with $\Delta_{\mathrm{c}}(t)=\Delta_{\mathrm{FF,TS}}(t)$ in Eq.~\eqref{eq:DeltacFFSt1}.
The fidelity of the control at the final time is defined as
\begin{linenomath}
\begin{align}
    F(t_{\mathrm{f}})=\left|\Braket{\Psi(t_{\mathrm{f}})|-\alpha_1,-\alpha_2,\frac{\sum_{j=1,2}\alpha_jg_{j\mathrm{c}}}{\Delta_{\mathrm{f}}}}\right|^2.
    \label{eq:fidelity1}
\end{align}
\end{linenomath}
For comparison, we also calculate the fidelity for the case in which
\begin{linenomath}
\begin{align}
    \Delta_{\mathrm{c}}(t)=\Delta_{0,\mathrm{TS}}(t)=\Delta_{\mathrm{i}}\left\{1+\left(\frac{\Delta_{\mathrm{i}}}{\Delta_{\mathrm{f}}}-1\right)g[\Lambda_0(t)]\right\}^{-1}.\label{eq:Deltac0TS2}
\end{align}
\end{linenomath}
The infidelity $1-F(t_{\mathrm{f}})$ is shown in Fig.~\ref{fig:infidelity1} for various values of $t_{\mathrm{f}}$.
The parameter values used are listed in Table~\ref{table:parameter1}.
In the range $50$ ns $\leq t_{\mathrm{f}}\leq100$ ns, the infidelity when $\Delta_{\mathrm{c}}(t)=\Delta_{\mathrm{FF,TS}}(t)$ (the magenta diamonds) is almost the same with that when $\Delta_{\mathrm{c}}(t)=\Delta_{0,\mathrm{TS}}(t)$ (the green circles).
In this range, the effect of fast-forwarding is small.
By contrast, in the range 10.4~ns $\leq t_{\mathrm{f}}\leq45$ ns, the former infidelity is smaller than the latter.
The difference becomes larger as $t_{\mathrm{f}}$ becomes shorter.
At $t_{\mathrm{f}}=10.4$~ns (the shortest $t_{\mathrm{f}}$), the former is approximately 0.4\% of the latter and is the smallest of all the magenta diamonds.
This demonstrates that our method can realize high-fidelity displacement of the coupler in a very short time.

We note one limitation of the present protocol.
If the ramp time $t_{\mathrm f}$ is shorter than a
certain threshold (approximately $10.4$ ns in the
above simulation), there exists a time $t$ between
$0$ and $T_{\mathrm f}$ such that
\begin{linenomath}
\begin{align}
1+\left(\frac{\Delta_{\mathrm{i}}}{\Delta_{\mathrm{f}}}-1\right)\left[g(t)+\frac{60w(t)}{\Delta_{\mathrm{i}}^2T_{\mathrm{f}}^2}\right]=0,
\end{align}
\end{linenomath}
which makes $\Delta_{\mathrm c}(t)$ in
Eq.~\eqref{eq:DeltacFFSt1} diverge.
In this case, the required detuning trajectory no
longer exists and the protocol breaks down.

\begin{table}
    \caption{The parameter values of the system.
    For brevity, the two KPOs are made the same.
    Here, $j\in\{1,2\}$.
    \label{table:parameter1}}
    \begin{center}
      \begin{tabular}{lc}
      \hline\hline
      $K_j/2\pi$ (MHz) & $2$ \\
      $p_j/2\pi$ (MHz) & $8$ \\
      $\Delta_{\mathrm{i}}/2\pi$ (MHz) & $200$ \\
      $\Delta_{\mathrm{f}}/2\pi$ (MHz) & $20$ \\
      $g_{j\mathrm{c}}/2\pi$ (MHz) & $2$ \\
      $g_{12}/2\pi$ (kHz) & $20$ \\
      \hline\hline
      \end{tabular}
    \end{center}
\end{table}

The infidelity of our method in Fig.~\ref{fig:infidelity1} (the magenta diamonds) stems mainly from deviation of the two-KPO state from $|-\alpha_1,-\alpha_2\rangle$ due to $\hat{H}_{ZZ}(t)$ in Eq.~\eqref{eq:HZZt}.
To see this, we consider the following Hamiltonian of the two KPOs:
\begin{linenomath}
\begin{align}
    \hat{H}_{2\mathrm{KPOs}}&=\hbar\sum_{j=1,2}\left[-\frac{K_j}{2}(\hat{a}_j^{\dag2}-\alpha_j^2)(\hat{a}_j^{2}-\alpha_j^2)+\frac{K_j\alpha_j^4}{2}\right]\notag \\
    &\quad\mbox{}+\hat{H}_{ZZ}(t).
\end{align}
\end{linenomath}
We prepare the initial state $|\tilde{\Psi}(0)\rangle=|-\alpha_1,-\alpha_2\rangle$, time evolve it under $\hat{H}_{2\mathrm{KPOs}}$ until $t=t_{\mathrm{f}}$, and calculate the fidelity
\begin{linenomath}
\begin{align}
    \tilde{F}(t_{\mathrm{f}})=\left|\Braket{\tilde{\Psi}(t_{\mathrm{f}})|-\alpha_1,-\alpha_2}\right|^2.
    \label{eq:fidelity2}
\end{align}
\end{linenomath}
The infidelity $1-\tilde{F}(t_{\mathrm{f}})$ when $\Delta_{\mathrm{c}}(t)=\Delta_{\mathrm{FF,TS}}(t)$ (the blue pentagons) and when $\Delta_{\mathrm{c}}(t)=\Delta_{0,\mathrm{TS}}(t)$ (the red triangles) is also shown in Fig.~\ref{fig:infidelity1}.
The relative errors between the magenta diamonds and the blue pentagons are below 47\%.
Thus, we ascribe the infidelity $1-F(t_{\mathrm{f}})$ when $\Delta_{\mathrm{c}}(t)=\Delta_{\mathrm{FF,TS}}(t)$ primarily to leakage of the two-KPO state out of $|-\alpha_1,-\alpha_2\rangle$ due to $\hat{H}_{ZZ}(t)$;
$\hat{a}^{\dag}_j$ ($j\in\{1,2\}$) acts on $|-\alpha_j\rangle$ as
\begin{linenomath}
\begin{align}
    \hat{a}^{\dag}_j|-\alpha_j\rangle=\hat{D}(-\alpha_j)|1\rangle-\alpha_j|-\alpha_j\rangle,
    \label{eq:excitation1}
\end{align}
\end{linenomath}
causing the excitation to $\hat{D}(-\alpha_j)|1\rangle$.
This leakage deviates the effective Hamiltonian of the coupler from Eq.~\eqref{eq:Hceff11}, which makes the coupler displacement imperfect.
We suspect that this also causes the infidelity of our method.

We note several additional caveats that may
be relevant in practical implementations.

Achieving short ramp times requires large variations
of the coupler detuning, which may place stringent
demands on the bandwidth of the flux delivery line.
The spectrum of the coupled system may also restrict
the allowable range of the coupler detuning in order
to avoid unwanted resonances and to keep the system
close to the subspace spanned by the four states in
Eqs.~\eqref{eq:psiFF00t}--\eqref{eq:psiFF11t}.

Furthermore, reducing the ramp time increases the
excursion of the coupler in the momentum quadrature,
which can enhance the effects of the coupler
self-Kerr nonlinearity, cross-Kerr-induced detuning
shifts, and magnetic-flux noise; see
Appendix~\ref{sec:self}.
The larger excursion may also increase leakage
outside the computational subspace.

For the parameter range considered in Fig.~\ref{fig:infidelity1}, however,
the threshold $t_{\mathrm f}\approx10.4$ ns is
reached while the momentum excursion remains
modest, satisfying $\max_{0\leq t\leq t_{\mathrm{f}}}\dot{\alpha}_0[\Lambda(t)]/\Delta_{\mathrm{i}}<0.3$
even at $t_{\mathrm f}=10.4$ ns.
Therefore, although the above effects constitute
important caveats, our numerical results do not
suggest that they become the factor determining the
minimum ramp time accessible within the present
protocol.

\begin{figure}
    \centering
    \includegraphics[width=0.46\textwidth]{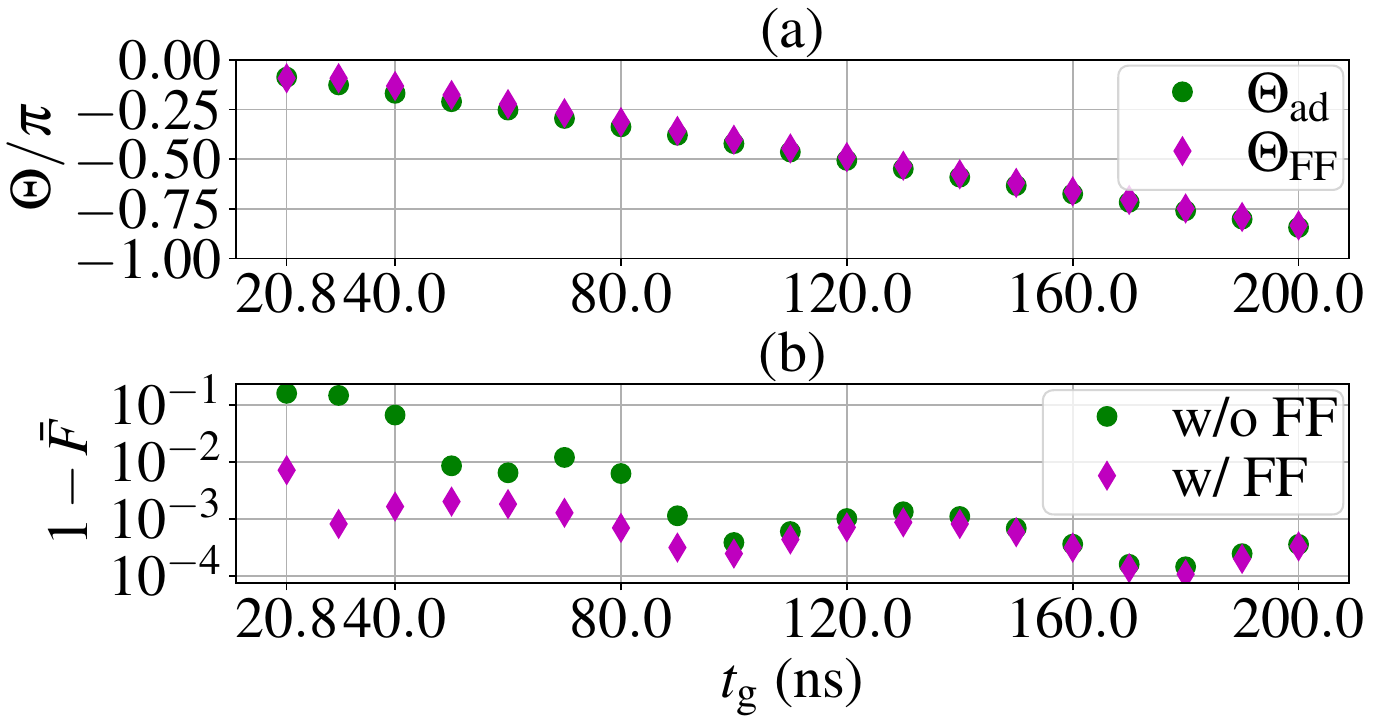}
    \caption{(a) $\Theta_{\mathrm{ad}}$ [Eq.~\eqref{eq:Thetaad1}] when $\Delta_{\mathrm{c}}(t)=\Delta_{\mathrm{0,TS}}(t)$ [Eq.~\eqref{eq:Deltac0TS2}] for $0\leq t\leq t_{\mathrm{g}}/2$ and $\Delta_{\mathrm{c}}(t)=\Delta_{\mathrm{0,TS}}(t_{\mathrm{g}}-t)$ for $t_{\mathrm{g}}/2\leq t\leq t_{\mathrm{g}}$; $\Theta_{\mathrm{FF}}$ [Eq.~\eqref{eq:ThetaFF}] when $\Delta_{\mathrm{c}}(t)=\Delta_{\mathrm{FF,TS}}(t)$ [Eq.~\eqref{eq:DeltacFFSt1}] for $0\leq t\leq t_{\mathrm{g}}/2$ and $\Delta_{\mathrm{c}}(t)=\Delta_{\mathrm{FF,TS}}(t_{\mathrm{g}}-t)$ for $t_{\mathrm{g}}/2\leq t\leq t_{\mathrm{g}}$.
    (b) The average gate infidelity $1-\bar{F}$ [Eq.~\eqref{eq:ave_fidelity1}] with and without fast-forward.
    The parameter values used are listed in Table~\ref{table:parameter1}.
    \label{fig:ave_infidelity1}}
\end{figure}
We also calculate the average fidelity of the $\hat{R}_{\mathrm{ZZ}}$ gate~\cite{PEDERSEN200747},
\begin{linenomath}
\begin{align}
    \bar{F}=\frac{\mathrm{Tr}(\hat{M}\hat{M}^{\dag})+|\mathrm{Tr}(\hat{M})|^2}{20}, \label{eq:ave_fidelity1}
\end{align}
\end{linenomath}
where $\hat{M}=\hat{U}_0^{\dag}\hat{U}$ with $\hat{U}_0=\hat{R}_{\mathrm{ZZ}}(\Theta)$ and
\begin{linenomath}
\begin{align}
    \hat{U}&=\sum_{k,l,m,n=0,1}\left\langle\widetilde{k,l}\left|\exp\left(-\frac{\mathrm{i}}{\hbar}\int_0^{t_{\mathrm{g}}}\hat{H}(t)\,\mathrm{d}t\right)\right|\widetilde{m,n}\right\rangle \notag \\
    &\quad\times|\widetilde{k,l}\rangle\langle\widetilde{m,n}|.
\end{align}
\end{linenomath}
For the case with fast-forward, we set $\Delta_{\mathrm{c}}(t)=\Delta_{\mathrm{FF,TS}}(t)$ [Eq.~\eqref{eq:DeltacFFSt1}] for $0\leq t\leq t_{\mathrm{g}}/2$, $\Delta_{\mathrm{c}}(t)=\Delta_{\mathrm{FF,TS}}(t_{\mathrm{g}}-t)$ for $t_{\mathrm{g}}/2\leq t\leq t_{\mathrm{g}}$, and $\Theta=\Theta_{\mathrm{FF}}$ [Eq.~\eqref{eq:ThetaFF}].
For the case without fast-forward, we set $\Delta_{\mathrm{c}}(t)=\Delta_{\mathrm{0,TS}}(t)$ [Eq.~\eqref{eq:Deltac0TS2}] for $0\leq t\leq t_{\mathrm{g}}/2$, $\Delta_{\mathrm{c}}(t)=\Delta_{\mathrm{0,TS}}(t_{\mathrm{g}}-t)$ for $t_{\mathrm{g}}/2\leq t\leq t_{\mathrm{g}}$, and $\Theta=\Theta_{\mathrm{ad}}$ [Eq.~\eqref{eq:Thetaad1}].
The parameter values used are the same as those in Fig.~\ref{fig:infidelity1}; see Table~\ref{table:parameter1}.
$\Theta_{\mathrm{FF}}$, $\Theta_{\mathrm{ad}}$, and $1-\bar{F}$ for various values of $t_{\mathrm{g}}$ are shown in Fig.~\ref{fig:ave_infidelity1}.
The average gate infidelity with fast-forward is smaller than that without fast-forward and than 0.8\% for $20.8$~ns $\leq t_{\mathrm{g}}\leq200$ ns.

\section{Conclusions}
\label{sec:Conclusions}
In this work, we applied fast-forward scaling theory to the tunable-coupler-based $R_{ZZ}$ gate for Kerr-cat qubits.
To this end, we developed a protocol for fast and high-fidelity displacement of the coupler through detuning modulation and applied it to the tunable-coupler-based gate scheme.
As a result, we realized a high-speed $R_{ZZ}$ gate while suppressing nonadiabatic excitation of the coupler.

Numerical simulations showed that, for a ramp time of $t_{\mathrm f}=10.4$ ns, the displacement infidelity with fast-forward scaling is reduced to approximately $0.4\%$ of that without fast-forward scaling in the ideal case. These results demonstrate that fast-forward scaling provides an effective route to accelerating tunable-coupler-based entangling gates in Kerr-cat-qubit architectures.

\begin{acknowledgments}
This paper is partly based on results obtained from a project, JPNP16007, commissioned by the New Energy and Industrial Technology Development Organization (NEDO), Japan.
S.M. acknowledges the support from JST [Moonshot R\&D] [Grant Number JPMJMS2061].
\end{acknowledgments}

\appendix

\section{Derivation of the adiabatic \texorpdfstring{$R_{ZZ}$}{RZZ} gate}
\label{app:RZZ}
In this Appendix, we derive the rotation angle
$\Theta_{\mathrm{ad}}$ of the adiabatic $R_{ZZ}$ gate
given in Eq.~\eqref{eq:Thetaad1}.

We tune $\Delta_{\mathrm c}(t)$ sufficiently slowly so that the state remains in the subspace spanned by
$\{
\ket{\psi_{0,0}(t)},
\ket{\psi_{0,1}(t)},
\ket{\psi_{1,0}(t)},
\ket{\psi_{1,1}(t)}
\}$
and $\hat H_{ZZ}(t)$ can be treated as a perturbation.

The corresponding eigenenergies to first order in perturbation theory,
\begin{linenomath}
\begin{align}
    \{E_{k,l}(t):=\langle\psi_{k,l}(t)|\hat{H}(t)|\psi_{k,l}(t)\rangle|k,l=0,1\},
\end{align}
\end{linenomath}
are 
\begin{linenomath}
\begin{align}
    E_{0,0}(t)&=E_{1,1}(t)=E^{(0)}+E^{(1)}(t), \label{eq:E00E11}\\
    E_{0,1}(t)&=E_{1,0}(t)=E^{(0)}-E^{(1)}(t). \label{eq:E01E10}
\end{align}
\end{linenomath}

We prepare the initial state of the system as
\begin{linenomath}
\begin{align}
    \ket{\Psi(0)}=\sum_{k,l=0}^1\beta_{k,l}\ket{\widetilde{k,l}},
    \label{eq:app_initial1}
\end{align}
\end{linenomath}
where $\{\beta_{k,l}\}$ are coefficients.

Under adiabatic evolution satisfying Eq.~\eqref{eq:noRX1}, each computational basis state acquires only a
dynamical phase. Therefore, the state at $t=t_{\mathrm g}$ is approximately given by
\begin{linenomath}
\begin{align}
    \ket{\Psi(t_{\mathrm{g}})}&=\mathcal{T}\exp\left(
        -\frac{\mathrm{i}}{\hbar}\int_0^{t_{\mathrm{g}}}\hat{H}(t)\,\mathrm{d}t
    \right)\ket{\Psi(0)} \notag \\
    &\approx\mathrm{e}^{-\mathrm{i}E^{(0)}t_{\mathrm{g}}/\hbar}\hat{R}_{ZZ}(\Theta_{\mathrm{ad}})\ket{\Psi(0)}=:\ket{\Psi_{\mathrm{ad}}(t_{\mathrm{g}})},
\end{align}
\end{linenomath}
where $\mathcal{T}$ is the time-ordering operator.
The operator
\begin{linenomath} 
\begin{align}
    \hat{R}_{ZZ}(\Theta):=\sum_{k,l=0}^1\mathrm{e}^{-\mathrm{i}(2\delta_{k,l}-1)\Theta/2}\ket{\widetilde{k,l}}\bra{\widetilde{k,l}}
\end{align}
\end{linenomath}
is the $R_{ZZ}$-gate operator with rotation angle $\Theta$, and
\begin{linenomath}
\begin{align}
    \Theta_{\mathrm{ad}}:=2\int_0^{t_{\mathrm{g}}}\frac{E^{(1)}(t)}{\hbar}\,\mathrm{d}t. 
\end{align}
\end{linenomath}

Although a resonant drive can perfectly displace a resonator in a very short time regardless of adiabaticity as shown in Appendix~\ref{app:resonant}, the following constraint on the coupler mode rejects a resonant-drive strategy;  the state of the coupler at $t=t_{\mathrm{g}}$ after an $R_{ZZ}$ gate must be $|-\alpha_{\mathrm{c}}^{+}(t_{\mathrm{g}})\rangle=|-\alpha_{\mathrm{c}}^{+}(0)\rangle$ and $|\alpha_{\mathrm{c}}^{+}(t_{\mathrm{g}})\rangle=|\alpha_{\mathrm{c}}^{+}(0)\rangle$ when the input computational state is $|\widetilde{0,0}\rangle$ and $|\widetilde{1,1}\rangle$, respectively.
This is because the state of the system must remain in the computational subspace after an $R_{ZZ}$ gate.
If we set $\Delta_{\mathrm{c}}(t)=0$ for $0\leq t\leq t_{\mathrm{g}}$, the coupler will be displaced to undesired directions, and the state of the system will leave the computational subspace; see Appendix~\ref{sec:resonant}.

\section{Perfect displacement of a resonator under a resonant drive}
\label{app:resonant}
Let us consider a superconducting resonator with resonance frequency $\omega$ and negligible anharmonicity under a coherent drive with frequency $\omega_{\mathrm{d}}$ and complex amplitude $\Omega$.
The Hamiltonian in the laboratory frame is written as
\begin{linenomath}
\begin{align}
    \hat{H}_{\mathrm{lab}}/\hbar=\omega\hat{a}^\dagger \hat{a} - \Omega\mathrm{e}^{-\mathrm{i}\omega_{\mathrm{d}}t}\hat{a}^\dagger - \Omega^{*}\mathrm{e}^{\mathrm{i}\omega_{\mathrm{d}}t}\hat{a},
\end{align}
\end{linenomath}
where $\hbar$ is the reduced Planck constant; $\hat{a}$ is the annihilation operator.
In a rotating frame at frequency $\omega_{\mathrm{d}}$, the Hamiltonian is transformed into
\begin{linenomath}
\begin{align}
    \frac{\hat{H}}{\hbar} = \hat{U}_{\mathrm{rot}}\frac{\hat{H}_{\mathrm{lab}}}{\hbar}\hat{U}_{\mathrm{rot}}^{\dag}-\mathrm{i}\hat{U}_{\mathrm{rot}}\frac{\mathrm{d}}{\mathrm{d}t}\hat{U}_{\mathrm{rot}}^{\dag}
    =\Delta\hat{a}^\dagger \hat{a} - \Omega\hat{a}^\dagger - \Omega^{*}\hat{a},
\end{align}
\end{linenomath}
where $\hat{U}_{\mathrm{rot}}=\exp(\mathrm{i}\omega_{\mathrm{d}}\hat{a}^\dagger\hat{a}t)$ and $\Delta:=\omega-\omega_{\mathrm{d}}$ is a detuning.
When the drive is resonant ($\Delta=0$) and time-dependent, the time-evolution operator is given by $\hat{U}_{\mathrm{evo}}(t)=\mathrm{e}^{\mathrm{i}\eta(t)}\hat{D}[\gamma(t)]$, where $\hat{D}[\gamma(t)]=\exp[\gamma(t)\hat{a}^\dagger-\gamma^{*}(t)\hat{a}]$ with $\gamma(t)=\mathrm{i}\int_0^t\mathrm{d}t_1\Omega(t_1)$ is a displacement operator and $\eta(t)=\int_0^t\mathrm{d}t_1\int_0^{t_1}\mathrm{d}t_2\mathrm{Im}\left[\Omega(t_1)\Omega^{*}(t_2)\right]$ is a global phase~\cite[Sec.~3]{10.1063/1.1726550}.
In this case, high-speed perfect displacement is achievable with an appropriate $\Omega$ regardless of adiabaticity.

\section{The case where \texorpdfstring{$\Delta_{\mathrm{c}}(t)=0$}{Deltac0} for \texorpdfstring{$0\leq t\leq t_{\mathrm{g}}$}{tgrange}}
\label{sec:resonant}
If we set $\Delta_{\mathrm{c}}(t)=0$ for $0\leq t\leq t_{\mathrm{g}}$, the coupler will be displaced to undesired directions, and the state of the system will leave the computational subspace, as shown below.
For $t\leq0$, to suppress residual $ZZ$ coupling, we set $\Delta_{\mathrm{c}}(t)=g_{1\mathrm{c}}g_{2\mathrm{c}}/g_{12}$, which leads to $\alpha_{\mathrm{c}}^{+}(t)=2g_{12}\alpha_1/g_{2\mathrm{c}}$ ($g_{1\mathrm{c}}\alpha_1=g_{2\mathrm{c}}\alpha_2$); the state of the system lies in the subspace spanned by the following four computational states
\begin{linenomath}
\begin{align}
    |\widetilde{0,0}\rangle&=|\alpha_1,\alpha_2,-2g_{12}\alpha_1/g_{2\mathrm{c}}\rangle, \\
    |\widetilde{0,1}\rangle&=|\alpha_1,-\alpha_2,0\rangle, \\
    |\widetilde{1,0}\rangle&=|-\alpha_1,\alpha_2,0\rangle, \\
    |\widetilde{1,1}\rangle&=|-\alpha_1,-\alpha_2,2g_{12}\alpha_1/g_{2\mathrm{c}}\rangle.
\end{align}
\end{linenomath}
After $t=t_{\mathrm{g}}$, we want the state of the system to lie in the subspace, too.
For $0\leq t\leq t_{\mathrm{g}}$, since $\Delta_{\mathrm{c}}(t)=0$, the Hamiltonian $\hat{H}(t)$ is
\begin{linenomath}
\begin{align}
    \hat{H}(t)&=\sum_{j=1,2}\hat{H}_j(t)+\hat{H}_{\mathrm{I1}}+\hat{H}_{\mathrm{I2}}, \\
    \hat{H}_j(t)/\hbar&=-\frac{K_j}{2}\hat{a}_j^{\dag2}\hat{a}_j^2+\frac{p_j}{2}(\hat{a}_j^{\dag2}+\hat{a}_j^2), \\
    \hat{H}_{\mathrm{I1}}/\hbar&=g_{12}(\hat{a}_1^{\dag}\hat{a}_2+\hat{a}_1\hat{a}_2^{\dag}), \\
    \hat{H}_{\mathrm{I2}}/\hbar&=\sum_{j=1,2}g_{j\mathrm{c}}(\hat{a}_j^{\dag}\hat{a}_{\mathrm{c}}+\hat{a}_j\hat{a}_{\mathrm{c}}^{\dag}),
\end{align}
\end{linenomath}
where we also set $\Delta_{1}(t)=\Delta_{2}(t)=0$ to avoid unnecessary $R_X$ gates.
$\hat{H}_{\mathrm{I1}}$ contributes to an $R_{ZZ}$ gate, the speed of which depends on $g_{12}$.
$\hat{H}_{\mathrm{I2}}$ acts as an unwanted coherent drive on the coupler when the two KPOs are in phase.
For example, when the input state is $|\widetilde{0,0}\rangle$, the effective Hamiltonian of the coupler is given by
\begin{linenomath}
\begin{align}
    \hat{H}_{\mathrm{c},0,0}^{\mathrm{eff}}/\hbar=\langle\alpha_1,\alpha_2|\hat{H}_{\mathrm{I2}}/\hbar|\alpha_1,\alpha_2\rangle
    =2g_{1\mathrm{c}}\alpha_1(\hat{a}_{\mathrm{c}}+\hat{a}_{\mathrm{c}}^{\dag}),
\end{align}
\end{linenomath}
which leads to the following displacement operator on the coupler:
\begin{linenomath}
\begin{align}
    \mathrm{e}^{-\mathrm{i}\hat{H}_{\mathrm{c}}^{\mathrm{eff}}t/\hbar}
    =\exp[-2\mathrm{i}g_{1\mathrm{c}}\alpha_1t(\hat{a}_{\mathrm{c}}+\hat{a}_{\mathrm{c}}^{\dag})]
    =\hat{D}(-2\mathrm{i}g_{1\mathrm{c}}\alpha_1t).
\end{align}
\end{linenomath}
Thus, at $t=t_{\mathrm{g}}$, the state of the coupler is displaced by $-2\mathrm{i}g_{1\mathrm{c}}\alpha_1t_{\mathrm{g}}$ unnecessarily.
Similarly, when the input state is $|\widetilde{1,1}\rangle$, the state of the coupler is displaced by $2\mathrm{i}g_{1\mathrm{c}}\alpha_1t_{\mathrm{g}}$ unnecessarily.
Even if we have an external resonant drive, we cannot cancel both of the unnecessary displacements.
In this way, the state of the system leaves the computational subspace.

\section{Derivation of \texorpdfstring{$\ket{\psi_{\mathrm{FF}}(t)}$}{psiFFt} in Eq.~\eqref{eq:PsiFFt2}}
\label{app:derivation}
We derive $\ket{\psi_{\mathrm{FF}}(t)}$ in Eq.~\eqref{eq:PsiFFt2} from $\hat{H}_{\mathrm{FF}}(t)$ in Eq.~\eqref{HFF_drive_6_10_25} with the boundary conditions in Eqs.~\eqref{eq:alpha_boundary1} and the initial state $\ket{\psi_{\mathrm{FF}}(0)}=\ket{\psi_{\mathrm{ad}}(0)}$ in Eq.~\eqref{eq:Psiadt2}.
We have consulted Appendix A in Ref.~\cite{PhysRevA.83.013415}.
The Schr\"{o}dinger equation under $\hat{H}_{\mathrm{FF}}(t)$ is
\begin{linenomath}
\begin{align}
    \mathrm{i}\hbar\frac{\mathrm{d}}{\mathrm{d}t}\ket{\psi_{\mathrm{FF}}(t)}=\hat{H}_{\mathrm{FF}}(t)\ket{\psi_{\mathrm{FF}}(t)}.
\end{align}
\end{linenomath}
To seek a more tractable Hamiltonian, we consider the dynamics of $\ket{\tilde{\psi}_{\mathrm{FF}}(t)}=\hat{U}(t)\ket{\psi_{\mathrm{FF}}(t)}$, where $\hat{U}(t)$ is a unitary operator determined below.
We obtain
\begin{linenomath}
\begin{align}
    \mathrm{i}\hbar\frac{\mathrm{d}}{\mathrm{d}t}\ket{\tilde{\psi}_{\mathrm{FF}}(t)}=\tilde{H}_{\mathrm{FF}}(t)\ket{\tilde{\psi}_{\mathrm{FF}}(t)},
    \label{eq:SchrodingerFF1}
\end{align}
\end{linenomath}
where
\begin{linenomath}
\begin{align}
    \tilde{H}_{\mathrm{FF}}(t)=\hat{U}(t)\hat{H}_{\mathrm{FF}}(t)\hat{U}^{\dag}(t)+\mathrm{i}\hbar\left(\frac{\mathrm{d}}{\mathrm{d}t}\hat{U}(t)\right)\hat{U}^{\dag}(t).
    \label{eq:tildeHFF1}
\end{align}
\end{linenomath}
As the unitary operator, we use a displacement operator
\begin{linenomath}
\begin{align}
    \hat{U}(t)&=\hat{D}[\beta(t)]=\exp[\beta(t)\hat{a}^{\dag}-\beta^{*}(t)\hat{a}]\notag \\
    &=\mathrm{e}^{-|\beta(t)|^2/2}\mathrm{e}^{\beta(t)\hat{a}^{\dag}}\mathrm{e}^{-\beta^{*}(t)\hat{a}},
\end{align}
\end{linenomath}
which has the following properties:
\begin{linenomath}
\begin{gather}
    \hat{D}^{\dag}[\beta(t)]=\hat{D}[-\beta(t)], \\
    \hat{D}[\beta(t)]\hat{a}\hat{D}^{\dag}[\beta(t)]=\hat{a}-\beta(t).
\end{gather}
\end{linenomath}
We take into account classical-number terms in this Appendix.
Substituting $\hat{H}_{\mathrm{FF}}(t)$ in the last line of Eq.~\eqref{HFF_drive_6_10_25} into Eq.~\eqref{eq:tildeHFF1} leads to
\begin{widetext}
\begin{linenomath}
\begin{align}
    \tilde{H}_{\mathrm{FF}}(t)/\hbar&=\Delta\hat{a}^{\dag}\hat{a}
    -\Delta\left(\beta(t)-\mathrm{i}\frac{\dot{\beta}(t)}{\Delta}+\alpha_{\mathrm{FF}}(t)\right)\hat{a}^{\dag}
    -\Delta\left(\beta^{*}(t)-\mathrm{i}\frac{\dot{\beta}^{*}(t)}{\Delta}+\alpha_{\mathrm{FF}}(t)\right)\hat{a} \notag \\
    &\quad\mbox{}+\Delta[\beta(t)+\alpha_{\mathrm{FF}}(t)][\beta^{*}(t)+\alpha_{\mathrm{FF}}(t)]
    +\frac{\mathrm{i}}{2}[\beta(t)\dot{\beta}^{*}(t)-\beta^{*}(t)\dot{\beta}(t)].
\end{align}
\end{linenomath}
\end{widetext}
When
\begin{linenomath}
\begin{align}
    \beta(t)=-\alpha_0(t)-\mathrm{i}\frac{\dot{\alpha}_0(t)}{\Delta}=-\tilde{\alpha}(t),
\end{align}
\end{linenomath}
the Hamiltonian becomes simple:
\begin{linenomath}
\begin{align}
    \tilde{H}_{\mathrm{FF}}(t)/\hbar=\Delta\hat{a}^{\dag}\hat{a}+\alpha_{\mathrm{FF}}(t)\ddot{\alpha}_0(t)/\Delta.
\end{align}
\end{linenomath}
Then, giving the initial state
\begin{linenomath}
\begin{align}
    \ket{\tilde{\psi}_{\mathrm{FF}}(0)}&=\hat{U}(0)\ket{\psi_{\mathrm{FF}}(0)}=\hat{D}[\beta(0)]|\psi_{\rm ad}(0)\rangle =\sum_n c_n|n\rangle
\end{align}
\end{linenomath}
to the Schr\"{o}dinger equation \eqref{eq:SchrodingerFF1} returns the solution
\begin{linenomath}
\begin{align}
    \ket{\tilde{\psi}_{\mathrm{FF}}(t)}&=\exp\left(-\mathrm{i}\int_0^t\frac{\alpha_{\mathrm{FF}}(s)\ddot{\alpha}_0(s)}{\Delta}\mathrm{d}s\right) \notag \\
    &\quad\times\sum_n c_ne^{-\mathrm{i}n \Delta t}\ket{n}.
\end{align}
\end{linenomath}
Returning to the original frame, we arrive at
\begin{linenomath}
\begin{align}
    \ket{\psi_{\mathrm{FF}}(t)}&=\hat{D}^{\dag}[\beta(t)]\ket{\tilde{\psi}_{\mathrm{FF}}(t)} \notag \\
    &=\exp\left(-\mathrm{i}\int_0^t\frac{\alpha_{\mathrm{FF}}(s)\ddot{\alpha}_0(s)}{\Delta}\mathrm{d}s\right) \notag \\
    &\quad\times\hat{D}[\tilde{\alpha}(t)]\sum_n c_n e^{-\mathrm{i}n \Delta t}|n\rangle,
    \label{eq:psiFFt_gphase1}
\end{align}
\end{linenomath}
which is the same as Eq.~\eqref{eq:PsiFFt2} up to the global phase.

\section{Counterdiabatic method}
\label{sec:CD}
The unitary operator $\hat{U}_{\mathrm{CD}}(t)$ necessary to obtain $\ket{\psi_{\mathrm{ad}}(t)}$ in Eq.~\eqref{eq:Psiadt2} starting from $\ket{\psi_{\mathrm{ad}}(0)}$, that is, $\ket{\psi_{\mathrm{ad}}(t)}=\hat{U}_{\mathrm{CD}}(t)\ket{\psi_{\mathrm{ad}}(0)}$, is written as
\begin{linenomath}
\begin{align}
    \hat{U}_{\mathrm{CD}}(t)=\hat{D}[\alpha_0(t)]\sum_n e^{-\mathrm{i}n \Delta t}\ket{n}\bra{n}\hat{D}^{\dag}[\alpha_0(0)].
\end{align}
\end{linenomath}
The corresponding counterdiabatic Hamiltonian $\hat{H}_{\mathrm{CD}}(t)$, which satisfies
\begin{linenomath}
\begin{align}
    \mathrm{i}\hbar\frac{\mathrm{d}}{\mathrm{d}t}\ket{\psi_{\mathrm{ad}}(t)}=\hat{H}_{\mathrm{CD}}(t)\ket{\psi_{\mathrm{ad}}(t)},
\end{align}
\end{linenomath}
is derived straightforwardly as
\begin{linenomath}
\begin{align}
    \hat{H}_{\mathrm{CD}}(t)&=\mathrm{i}\hbar\frac{\mathrm{d}\hat{U}_{\mathrm{CD}}(t)}{\mathrm{d}t}[\hat{U}_{\mathrm{CD}}(t)]^{-1}
    \notag \\
    &=\hat{H}_0(t)+\mathrm{i}\hbar\dot{\alpha}_0(t)(\hat{a}^{\dag}-\hat{a})
    \notag \\
    &=\hbar\Delta\hat{D}[\alpha_{\mathrm{CD}}(t)]\hat{a}^{\dag}\hat{a}\hat{D}^{\dag}[\alpha_{\mathrm{CD}}(t)],
\end{align}
\end{linenomath}
where $\hat{H}_0(t)$ is given in Eq.~\eqref{eq:H0t2} and
\begin{linenomath}
\begin{align}
    \alpha_{\mathrm{CD}}(t)=\alpha_0(t)-\mathrm{i}\dot{\alpha}_0(t)/\Delta.
\end{align}
\end{linenomath}

\section{Effects of the self-Kerr nonlinearity of the coupler, cross-Kerr interactions, and magnetic flux noise in the coupler}
\label{sec:self}
When the self-Kerr term,
\begin{linenomath}
\begin{align}
    \hat{H}_{\mathrm{c,Kerr}}/\hbar = -\frac{K_{\mathrm{c}}}{2}\hat{a}_{\mathrm{c}}^{\dag2}\hat{a}_{\mathrm{c}}^2,
\end{align}
\end{linenomath}
can be treated as a perturbation to $\hat{H}_{0\mathrm{th}}(t)$, it contributes to the $ZZ$ coupling~\cite{10.1063/5.0241315}, because
\begin{linenomath}
\begin{align}
    \langle\psi_{k,l}^{\mathrm{FF}}(t)|\hat{H}_{\mathrm{c,Kerr}}/\hbar|\psi_{k,l}^{\mathrm{FF}}(t)\rangle=0
\end{align}
\end{linenomath}
for $k\neq l$ ($k,l\in\{0,1\}$), $0\leq t\leq t_{\mathrm{g}}$ and
\begin{linenomath}
\begin{align}
    \langle\psi_{k,k}^{\mathrm{FF}}(t)|\hat{H}_{\mathrm{c,Kerr}}/\hbar|\psi_{k,k}^{\mathrm{FF}}(t)\rangle=-\frac{K_{\mathrm{c}}}{2}\{\tilde{\alpha}[\Lambda(t)]\}^4 
    \label{eq:self_Kerr_kk1}
\end{align}
\end{linenomath}
for $0\leq t\leq t_{\mathrm{g}}/2$. For $t_{\mathrm{g}}/2\leq t\leq t_{\mathrm{g}}$, $\Lambda(t)$ in Eq.~\eqref{eq:self_Kerr_kk1} is replaced by $\Lambda(t_{\mathrm{g}}-t)$.
Meanwhile, because
\begin{linenomath}
\begin{align}
    &\hat{a}_{\mathrm{c}}^{\dag2}\hat{a}_{\mathrm{c}}^2|\pm\tilde{\alpha}[\Lambda(t)]\rangle=\{\tilde{\alpha}[\Lambda(t)]\}^2\Big(\sqrt{2}\hat{D}\{\pm\tilde{\alpha}[\Lambda(t)]\}|2\rangle \notag \\
    &\mbox{}\pm2\tilde{\alpha}[\Lambda(t)]\hat{D}\{\pm\tilde{\alpha}[\Lambda(t)]\}|1\rangle+\{\tilde{\alpha}[\Lambda(t)]\}^2|\pm\tilde{\alpha}[\Lambda(t)]\rangle\Big),
    \label{eq:self_Kerr_excite1}
\end{align}
\end{linenomath}
the self-Kerr term excites the coupler state $|\psi_{\mathrm{c},k,k}^{\mathrm{eff}}(t)\rangle$, which prevents the perfect displacement of the coupler.
Hence, $K_{\mathrm{c}}$ should be as small as possible.

The above also applies to the cross-Kerr term between the $j$th KPO and the coupler,
\begin{linenomath}
\begin{align}
    \hat{H}_{j\mathrm{c},\mathrm{Kerr}}/\hbar=\chi_{j\mathrm{c}}\hat{a}^{\dag}_j\hat{a}_j\hat{a}_{\mathrm{c}}^{\dag}\hat{a}_{\mathrm{c}},
    \label{eq:cross_Kerr1}
\end{align}
\end{linenomath}
because
\begin{linenomath}
\begin{align}
    \langle\psi_{k,l}^{\mathrm{FF}}(t)|\hat{H}_{j\mathrm{c,Kerr}}/\hbar|\psi_{k,l}^{\mathrm{FF}}(t)\rangle=0
\end{align}
\end{linenomath}
for $k\neq l$ ($k,l\in\{0,1\}$), $0\leq t\leq t_{\mathrm{g}}$,
\begin{linenomath}
\begin{align}
    \langle\psi_{k,k}^{\mathrm{FF}}(t)|\hat{H}_{j\mathrm{c,Kerr}}/\hbar|\psi_{k,k}^{\mathrm{FF}}(t)\rangle=\chi_{j\mathrm{c}}\alpha_j^2\{\tilde{\alpha}[\Lambda(t)]\}^2 
    \label{eq:cross_Kerr_kk1}
\end{align}
\end{linenomath}
for $0\leq t\leq t_{\mathrm{g}}/2$, and
\begin{linenomath}
\begin{align}
    &\quad\hat{a}_{\mathrm{c}}^{\dag}\hat{a}_{\mathrm{c}}|\pm\tilde{\alpha}[\Lambda(t)]\rangle\notag \\
    &=\pm\tilde{\alpha}[\Lambda(t)]\Big(\hat{D}\{\pm\tilde{\alpha}[\Lambda(t)]\}|1\rangle\pm\tilde{\alpha}[\Lambda(t)]|\pm\tilde{\alpha}[\Lambda(t)]\rangle\Big).
    \label{eq:acdag_ac1}
\end{align}
\end{linenomath}
The excitation in Eq.~\eqref{eq:acdag_ac1} is also caused by the pure dephasing of the coupler due to magnetic flux noise~\cite[Sec.~III.C.2]{doi:10.1063/1.5089550}.
Excitations of the coupler yield excitations of the KPOs through interaction.
Note that the cross-Kerr term directly excites the $j$th KPO.
The cross-Kerr term between the two KPOs,
\begin{linenomath}
\begin{align}
    \hat{H}_{12,\mathrm{Kerr}}/\hbar=\chi_{12}\hat{a}^{\dag}_1\hat{a}_1\hat{a}_{2}^{\dag}\hat{a}_{2},
\end{align}
\end{linenomath}
also excites them.
Excitations of the KPOs induce their bit flips, and weaken their noise biases~\cite{PhysRevLett.128.110502}.
The induced bit-flip errors can be suppressed by using colored (frequency-selective) dissipation~\cite{PhysRevLett.128.110502} or circuit refrigeration~\cite{Masuda2025}.

From another viewpoint, the cross-Kerr term in Eq.~\eqref{eq:cross_Kerr1} shifts the detuning of the coupler, because the term effectively acts as a harmonic term of the coupler.
For example, when the two-KPO state is $|-\alpha_1,-\alpha_2\rangle$, the effective Hamiltonian of the coupler is given by
\begin{linenomath}
\begin{align}
    &\quad\hat{H}_{\mathrm{c},1,1}^{\mathrm{eff,cross}}(t)
    =\hat{H}_{\mathrm{c},1,1}^{\mathrm{eff}}(t) \notag \\
    &\quad\mbox{}+\langle-\alpha_1,-\alpha_2|\hat{H}_{1\mathrm{c},\mathrm{Kerr}}+\hat{H}_{2\mathrm{c},\mathrm{Kerr}}|-\alpha_1,-\alpha_2\rangle \notag \\
    &=\hat{H}_{\mathrm{c},1,1}^{\mathrm{eff}}(t)
    +\sum_{j=1,2}\hbar\chi_{j\mathrm{c}}\alpha_j^2\hat{a}_{\mathrm{c}}^{\dag}\hat{a}_{\mathrm{c}},
    \label{eq:Heff_cross1}
\end{align}
\end{linenomath}
where $\hat{H}_{\mathrm{c},1,1}^{\mathrm{eff}}(t)$ is given in Eq.~\eqref{eq:Hceff11} with $\Delta_{\mathrm{c}}(t)=\Delta_{\mathrm{FF,TS}}(t)$ in Eq.~\eqref{eq:DeltacFFSt1}.

\begin{figure*}
    \centering
    \includegraphics[width=0.8\textwidth]{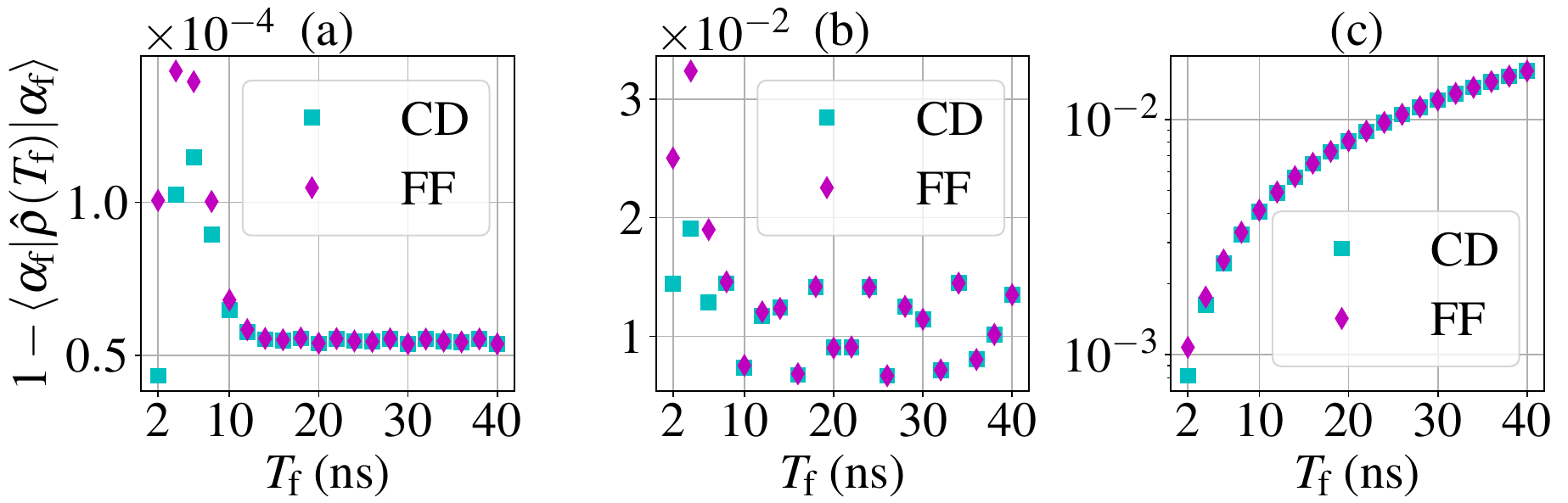}
    \caption{The infidelity $1-\langle\alpha_{\mathrm{f}}|\hat{\rho}(T_{\mathrm{f}})|\alpha_{\mathrm{f}}\rangle$ under FF/CD dynamics; $\hat{\rho}(T_{\mathrm{f}})=|\psi(T_{\mathrm{f}})\rangle\langle\psi(T_{\mathrm{f}})|$ in (a) and (b).
    The resonator state time evolves under the Hamiltonian \eqref{eq:FFCD_selfKerr1} with $K/2\pi=180$ kHz in (a), under the Hamiltonian \eqref{eq:FFCD_shift1} with $\Delta'/2\pi=-9.6$ MHz ($\chi_{1\mathrm{c}}=\chi_{2\mathrm{c}}=-1.2$ MHz) in (b), and under the GKSL equation \eqref{eq:pure_dephasing1} with $\kappa/2\pi=36$ kHz in (c).
The self-Kerr and cross-Kerr coefficients were
chosen to be comparable to experimentally reported
values for tunable resonators~\cite{PhysRevB.97.144502},
while the pure dephasing rate was chosen to be
comparable to experimentally reported values
~\cite{PhysRevApplied.20.014057}.
    The other parameter values used are the same with those used to calculate $1-|\braket{\psi_{\mathrm{nonad}}(T_{\mathrm{f}})|\alpha_{\mathrm{f}}}|^2$ in Fig.~\ref{fig:drive_infidelity1}.
    \label{fig:infidelity_self1} 
    }
\end{figure*}
As we can see from Eq.~\eqref{eq:tilde_alpha_lambda1} and Fig.~\ref{fig:alphat1}, the higher the gate speed is, the larger an excursion of the fast-forwarded trajectory in the momentum quadrature (Im[$\tilde{\alpha}[\Lambda(t)]$] in Fig.~\ref{fig:trajectories}) is.
To investigate whether or not an excursion in the momentum quadrature amplifies errors in resonator displacement due to self-Kerr nonlinearity, detuning shift, and pure dephasing, we compare the errors under FF dynamics with those under CD dynamics.
Note that the difference between the displacement under the FF dynamics, $\tilde{\alpha}(t)$, and that under the CD dynamics, $\alpha_0(t)$, is $\mathrm{i}\dot{\alpha}_0(t)/\Delta$; see Eq.~\eqref{eq:tilde_alpha1} and Fig.~\ref{fig:trajectories}.
To evaluate errors due to self-Kerr nonlinearity $K$ under FF/CD dynamics, we time evolve the resonator state $|\psi(t)\rangle$ under the following Hamiltonian:
\begin{linenomath}
\begin{align}
    \hat{H}_{\mathrm{FF/CD}}(t)-\frac{\hbar K}{2}\hat{a}^{\dag2}\hat{a}^2, \label{eq:FFCD_selfKerr1}
\end{align}
\end{linenomath}
where $\hat{H}_{\mathrm{FF/CD}}(t)$ is given in Eq.~\eqref{HFF_drive_6_10_25}/\eqref{eq:HCD1}, starting from the initial state $|\psi(0)\rangle=|\alpha_{\mathrm{i}}\rangle$ and calculate the infidelity between the final state $|\psi(T_{\mathrm{f}})\rangle$ and the target state $|\alpha_{\mathrm{f}}\rangle$.
Similarly, errors due to detuning shift $\Delta'$ are calculated using the following Hamiltonian:
\begin{linenomath}
\begin{align}
    \hat{H}_{\mathrm{FF/CD}}(t)+\hbar\Delta'\hat{a}^{\dag}\hat{a}
    \label{eq:FFCD_shift1}
\end{align}
\end{linenomath}
and those due to pure dephasing are calculated using the following Gorini–Kossakowski–Sudarshan–Lindblad (GKSL) equation~\cite{Gorini1976,Lindblad1976}:
\begin{linenomath}
\begin{align}
    \frac{\mathrm{d}\hat{\rho}(t)}{\mathrm{d}t}
    &=-\frac{\mathrm{i}}{\hbar}[\hat{H}_{\mathrm{FF/CD}}(t),\hat{\rho}(t)] \notag \\
    &\quad\mbox{}+\kappa
    \left(2\hat{a}^{\dag}\hat{a}\hat{\rho}(t)\hat{a}^{\dag}\hat{a}
    -\left\{(\hat{a}^{\dag}\hat{a})^2,
    \hat{\rho}(t)\right\}\right),
    \label{eq:pure_dephasing1}
\end{align}
\end{linenomath}
where $\kappa$ is the dephasing rate.
The calculation results are shown in Fig.~\ref{fig:infidelity_self1}.
When $T_{\mathrm{f}}$ is small, the infidelity under FF dynamics is larger than that under CD dynamics in each of the three cases.
Hence, an excursion in the momentum quadrature amplifies errors in resonator displacement due to self-Kerr nonlinearity, detuning shift, and pure dephasing.

\bibliography{5ref}
\end{document}